\documentclass[a4paper,fleqn,usenatbib]{mnras}
\usepackage{pdflscape}
\usepackage[T1]{fontenc}
\usepackage{ae,aecompl}
\usepackage{graphicx}	
\usepackage{amsmath}	
\usepackage{amssymb}	
\usepackage{appendix}
\title[X-ray study of a sample of FR0 radio galaxies]{X-ray study of a sample of FR0 radio galaxies: unveiling the nature of
the central engine}
\author[E. Torresi et al.]{
E. Torresi,$^{1,2}$\thanks{E-mail: torresi@iasfbo.inaf.it}
P. Grandi,$^{2}$
A. Capetti,$^{3}$
R.~D. Baldi,$^{4}$
G. Giovannini,$^{1,5}$
\\
$^{1}$Dipartimento di Astronomia, Universit\`a di Bologna, Via Gobetti 93/2, I-40129 Bologna, Italy\\
$^{2}$INAF--OAS Bologna, Area della Ricerca CNR, Via Gobetti 101, I-40129 Bologna, Italy \\
$^{3}$INAF--Osservatorio Astrofisico di Torino, Strada Osservatorio 20, I-10025, Pino Torinese, Italy\\
$^{4}$Department of Physics and Astronomy, The University of Southampton, SO17 1BJ, UK\\
$^{5}$INAF--IRA Bologna, Via Gobetti 101, I-40129 Bologna, Italy
}

\date{Accepted 2018 February 22. Received 2018 February 21; in original form 2017 July 4}

\pubyear{2018}

\begin{document}
\label{firstpage}
\pagerange{\pageref{firstpage}--\pageref{lastpage}}
\maketitle

\begin{abstract}
FR0s are compact radio sources that represent the bulk of the Radio-Loud (RL)
AGN population, but they are still poorly understood.  Pilot studies on these
sources have been already performed at radio and optical wavelengths: here we
present the first X-ray study of a sample of 19 FR0 radio galaxies selected from 
the SDSS/NVSS/FIRST sample of Best \& Heckman (2012), with redshift $\leq$ 0.15, radio 
size $\leq$ 10~kpc and optically classified as 
low-excitation galaxies (LEG).
The X-ray spectra are modeled with a power-law component absorbed by Galactic
column density with, in some cases, a contribution from thermal extended gas.
The X-ray photons are likely produced by the jet as attested by the observed
correlation between X-ray (2-10~keV) and radio (5~GHz) luminosities, similar to FRIs. The
estimated Eddington-scaled luminosities indicate a low accretion rate.
Overall, we find that the X-ray properties of FR0s are indistinguishable from
those of FRIs, thus adding another similarity between AGN
associated with compact and extended radio sources. 
A comparison between FR0s and low luminosity BL Lacs, rules out important beaming 
effects in the X-ray emission of the compact radio galaxies.  
FR0s have different X-ray properties with respect to young radio sources (e.g. GPS/CSS sources), 
generally characterized by higher X-ray luminosities and more complex spectra.
In conclusion, the paucity of extended radio emission in FR0s is probably related
to the intrinsic properties of their jets that prevent the formation of extended structures,
and/or to intermittent activity of their engines.

\end{abstract}

\begin{keywords}
galaxies: active -- galaxies: jets -- X-ray: galaxies.
\end{keywords}

\section{Introduction} 

The recent wide field surveys performed in the optical and radio
bands (e.g. SDSS\footnote{Sloan Digital Sky Survey (York et al. 2000)} and FIRST\footnote{Faint Images of the Radio Sky at Twenty centimetres survey (Becker et al. 1995; Helfand et al. 2015)})
showed that the population of radio sources
associated with active galactic nuclei (AGN) is dominated by objects in
which the radio emission is unresolved or barely resolved in the
FIRST images (e.g. Best et al. 2005; Baldi \& Capetti 2010; Baldi et al. 2018a): this implies that they have
typical sizes of less than $\sim$10~kpc. In contrast, radio
galaxies selected by high-flux limited low-frequency surveys such as the 3C (Edge et al. 1959),
the 3CR (Bennett 1962), the 4C (Pilkington et al. 1965) and the B2 (Colla et al. 1975; Fanti et al. 1978) often extend to
hundreds of kpc (FRI/FRII) and appear resolved at the angular
resolution provided by FIRST.
Baldi et al. (2010, 2018a)  showed that the radio
galaxies selected in the local Universe at 1.4~GHz, with similar
bolometric luminosities, span in a broad distribution of radio
luminosities and sizes, from compact to resolved with clear extended radio emission.
It is important to note that a clear dichotomy is not present among
sources selected from classical  low-frequency radio catalogues (B2, 3C, 4C) and
radio galaxies selected in the local Universe by surveys as the FIRST.\\
The lack of a clear difference in luminosity and size distribution 
requires to adopt an arbitrary angular size threshold, 
that furthermore corresponds to a different physical scale depending on distance.
Therefore, a precise definition of the population of compact radio sources suffers from several
observational difficulties, mainly because they are selected from
surveys (e.g. FIRST, NVSS\footnote{National Radio Astronomy Observatory (NRAO) Very Large Array (VLA) Sky Survey (Condon et al. 1998)} and AT20G\footnote{The Australia Telescope 20~GHz survey (Murphy et al. 2010)}) limited in flux, resolution and sensitivity. 

Ghisellini (2011) firstly described the compact sources studied by
Baldi \& Capetti (2009, 2010) as FR0s. The FR0 nomenclature was then followed by
Sadler et al. (2014)  ``as a convenient way
of linking the compact radio sources seen in nearby galaxies into the
canonical Fanaroff-Riley classification scheme."
However, Sadler et al. found a more diversified population, e.g. with significant 
contribution of high-excitation galaxies (HEG), with respect to Baldi \& Capetti (2010). These differences
are likely related to a substantial distinction in the luminosity functions of the two samples
considered: the Sadler et al. sources extend to a radio power $\sim$100 times higher at a 
radio frequency 10 times higher than the sample selected by Baldi \& Capetti. \\
It is also clear that compact radio sources are a very
heterogeneous population and they can be produced by AGN with widely
different multi-wavelength properties. For example, although most of
them are radio-loud AGN, radio-quiet galaxies often show compact
radio cores sometimes associated with pc/kpc scale emission (Ulvestad \& Ho 2001;
Nagar et al. 2005; Baldi et al. 2018b). Furthermore, the properties of their hosts and nuclei
differ depending on the frequency and flux threshold at which they
are selected.

Considering the difficulties in univocally defining the class described above, 
Baldi \& Capetti suggested to restrict the FR0 definition to a sub-population of
compact radio sources whose compactness is not due to relativistic effects
and which do not follow the correlation between total and
core radio power of classical FRI and FRII sources (Giovannini et al. 1988).
Indeed,  a source property useful to try a comparison and to select different
populations with different properties could be the core dominance. Giovannini
et al. (1988) discussed the core dominance properties for all sources from the
3CR and B2 catalogues with the only selection effect on the declination and
galactic latitude. A clear correlation between the core and total radio power
was found useful to constrain the source orientation and jet velocity. The
best fit linear regression of LogP$_{\rm c}$ versus LogP$_{\rm t}$ gives (see Giovannini et al. 2001):

\begin{equation}
\rm LogP_{c}=(7.6\pm1.1)+(0.62\pm0.04)~LogP_{t}
\end{equation} 

where P$_{\rm c}$ is the core radio power at 5~GHz and P$_{\rm t}$ is the total radio power at 408~MHz.

In a pilot program of high resolution ($\sim 0\farcs2$) radio imaging of a small sample of compact
sources,  Baldi et al. (2015, hereafter B15) defined as genuine FR0 
those sources that appear  unresolved, or slightly resolved, on a scale of 1-3~kpc in the radio maps, that are
located in red massive ($\sim$10$^{11}$~M$_{\odot}$) early-type galaxies with high black hole masses
(M$_{\rm BH}$$\geq$10$^{8}$M$_{\odot}$) and that are spectroscopically classified in the optical as low-excitation
galaxies (LEG)\footnote{LEG have generally weaker [OIII]-line
emission with respect to high-excitation galaxies (HEG) that show
[OIII]/H$\alpha$>0.2 and equivalent width of [OIII]>3~\AA. (Laing
et al. 1994; Jackson \& Rawlings 1997). More recent definitions have been provided by Kewley et al. (2006) on the basis of the
L$_{\rm [OIII]}$/$\sigma$$^{4}$ quantity and Buttiglione et al. (2010) 
on the basis of the Excitation Index (EI) defined as EI=log[OIII]/H$\beta$-1/3(log[NII]/H$\alpha$+log[SII]/H$\alpha$+log[OI]/H$\alpha$).
In particular, LEG sources are characterized by EI$\leq$0.95.}.
The sources of the B15 sample are highly core-dominated, since most of the emission 
detected at 5$\arcsec$ (FIRST resolution) is included within a compact region unresolved at 45$\arcsec$ (NVSS resolution): this turns out in a core-dominance higher by factor of $\sim$30 for FR0s with respect to FRIs of the 3CR catalog.

\noindent
Line luminosity is a robust proxy of the radiative power of the AGN
and, at least for the sources with similar multi-wavelength properties,
of the accretion rate. At a given line luminosity, FR0s are $\sim$100
less luminous than FRIs in total radio power.
Therefore, the compact radio galaxies studied by B15 are not simply unresolved sources, but they show a genuine lack of extended radio emission at large scales. 
Possible explanations
have been proposed, such as: (i) FR0s could be short-lived and/or recurrent
episodes of AGN activity, not long enough for radio jets to develop at large
scales (Sadler et al. 2014; Sadler 2016), or (ii) FR0s produce slow jets
experiencing instabilities and entrainment in the dense interstellar medium of
the host galaxy corona that causes their premature disruption (Bodo et
al. 2013; B15; Baldi et al. 2018a).

In this paper we present the first systematic X-ray study of a sample of
FR0 radio galaxies. 
Since the radio selection of compact radio galaxies carried out 
by Baldi \& Capetti (2010) and B15 turns out to correspond to an optical selection, 
we adopt these radio and spectro-photometric characteristics to define our 
FR0 class of low-excitation radio galaxies. 
This classification differs from the other FR classes,
not only for the radio morphology but also for specific
spectro-photometric characteristics. 
The key aim of our work is to investigate the central regions of FR0s
through X-rays in an effort to shed light on the nature of their central
engine. A comparison with the radio, optical and X-ray properties of the FRI
radio galaxies is also pursued to further explore  differences/similarities
between these two classes of sources. 
Since the FR0/FRI comparison is one of the main drivers of the present study,
this motivates our selection of only LEG spectroscopic types. Furthermore, this stricter
definition of FR0 enables us to restrict on a more homogeneous population of compact sources,
avoiding confusion with e.g. Seyfert-like objects or GPS/CSS sources (see Section 4.3).
Data were taken from the public archives of the X-ray satellites
currently on-flight (e.g. \emph{XMM-Newton}, \emph{Chandra},
\emph{Swift}). Most of the X-ray data of our sample are unpublished. 

Incidentally, we note that very recently a FR0 radio galaxy, i.e. Tol1326-379,
has been associated for the first time with a $\gamma$-ray source (Grandi,
Capetti \& Baldi 2016). Tol1326-379 shows a GeV luminosity typical of FRIs but
with a steeper $\gamma$-ray spectrum that can be related to intrinsic jet
properties or to a different viewing angle. For this source, a \emph{Swift} Target of Opportunity
(ToO) observation was performed during
the writing of the paper.

The paper is organized as follows: in Section 2 we define the sample. In
Section 3 we describe the observations, data reduction and spectral analysis,
while the results are discussed in Section 4. Notes on single sources and details of the X-ray analysis are reported in Appendix~A. 
The multi-wavelength properties of the FRI comparative sample are listed in Appendix~B.
Throughout the paper we use the following
cosmological parameters: H$_{0}$= 70~km$^{-1}$~s$^{-1}$~Mpc$^{-1}$,
$\Omega_{m}$=0.3, $\Omega_{\lambda}$=0.7 (Spergel et al. 2007). 


\begin{table*}
\caption{Log of the observations of the FR0 sample.}
\label{tab1}      
\centering          
\begin{tabular}{l l l l l l}    
\hline\hline       
SDSS  name                    &Telescope    &ObsID       & Exposure [ks]     &Offset ['] \\
\hline
J004150.47$-$091811.2         &Chandra      &15173        &42.5              &3.4\\
J010101.12$-$002444.4$^{*}$   &Chandra      &8259	      &16.8              &0.0 \\
J011515.78+001248.4$^{*}$     &XMM          &0404410201   &54.0              &0.095\\
J015127.10$-$083019.3         &Swift        &00036976004  &5.6               &0.894\\
J080624.94+172503.7           &Swift        &00085577001  &1.3               &0.337\\
J092405.30+141021.5           &Chandra      &11734        &30.1              &0.0\\
J093346.08+100909.0           &Swift	     &00036989002  &12.2		        &1.867\\
J094319.15+361452.1           &Swift  	     &00036997001  &5.4 		        &3.437\\
J104028.37+091057.1           &XMM          &0038540401   &24.0              &0.003\\
J114232.84+262919.9           &XMM          &0556560101   &32.9              &13.7\\
J115954.66+302726.9           &Swift        &00090129001  &3.4               &2.803\\
J122206.54+134455.9           &Swift        &00083911002  &1.3               &7.641\\
J125431.43+262040.6           &Chandra      &3074         &5.8               &2.0\\
Tol1326$-$379                 &Swift        &00034308001  &4.3               &0.0\\
J135908.74+280121.3           &Chandra      &12283        &10.1              &3.1\\
J153901.66+353046.0           &Swift        &00090113002  &2.8               &0.96\\
J160426.51+174431.1           &Chandra      &4996         &22.1              &2.5\\
J171522.97+572440.2           &Chandra      &4194         &47.9              &0.0\\
J235744.10$-$001029.9$^{*,a}$ &XMM          &--           &--                &-- \\
\hline\\
\multicolumn{6}{l}{$^{*}$ The source is already present in the FR0 sample of B15.} \\  
        \multicolumn{6}{l}{$^{a}$This source is part of the Third XMM-Newton Serendipitous Source Catalog,}\\
\multicolumn{6}{l}{Sixth Data Release (3XMM-DR6; Rosen et al. 2016).}                     
\end{tabular}
\end{table*}


\begin{figure}
\includegraphics[width=\columnwidth]{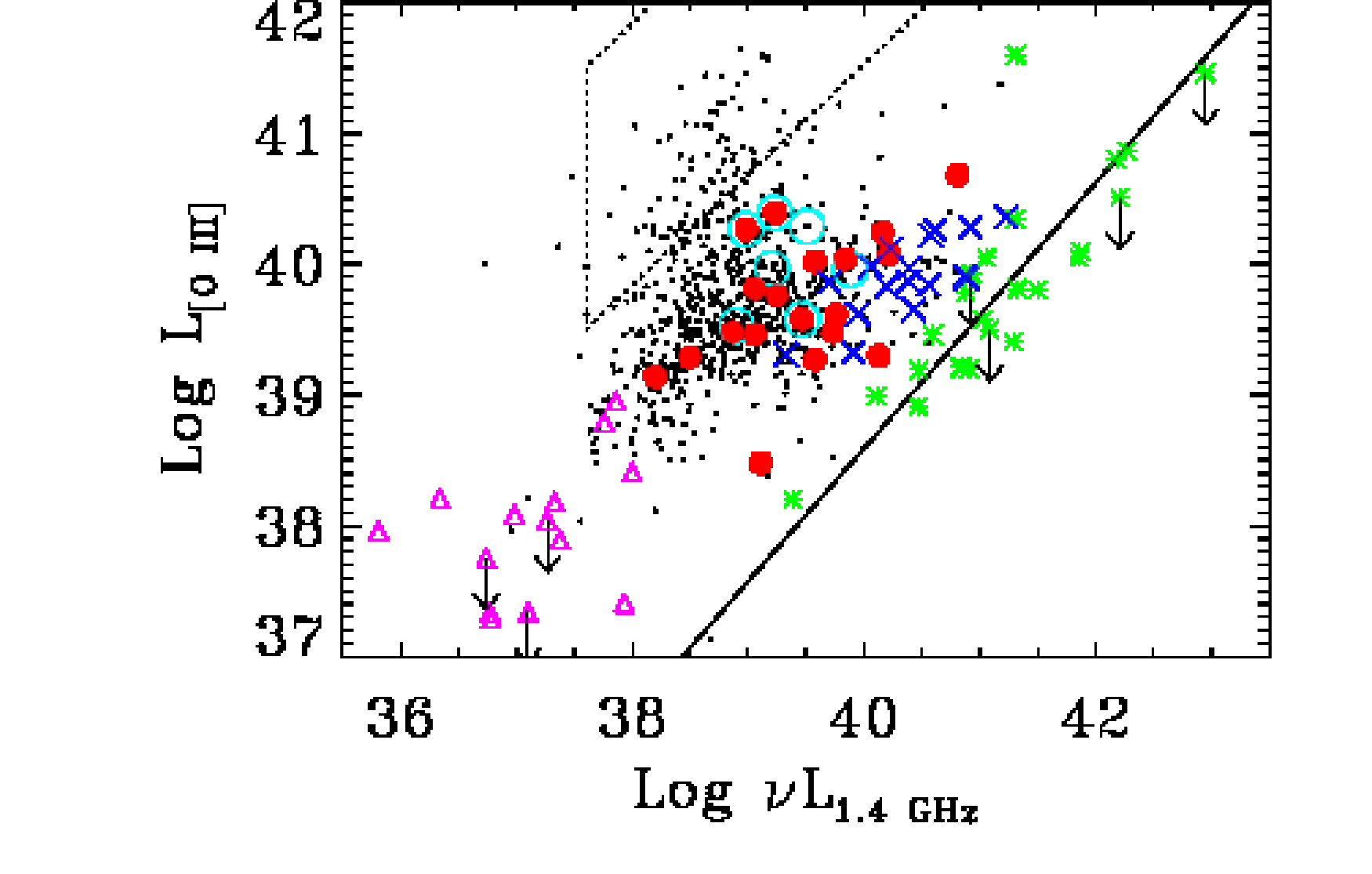}
\caption{FIRST versus [OIII] luminosities (both in erg~s$^{-1}$) adapted from B15. 
  {\it Red points} are the FR0s presented in this paper while {\it empty
    cyan circles} are the FR0s of B15. The {\it black dot points} correspond to the SDSS/NVSS
  sample analyzed by Baldi \& Capetti (2010), the {\it blue crosses} are the
  low-luminosity BL Lacs studied by Capetti \& Raiteri (2015), the {\it empty
    pink triangles} are the CoreG galaxies (Balmaverde \& Capetti 2006) and
  the {\it green stars} are the FRIs of the 3CR sample. The dashed line marks
  the boundary of the location of Seyfert galaxies. The solid line represents
  the line-radio correlation followed by the 3CR/FRIs.}
\label{fig1}
\end{figure}

\section{Sample selection}
In order to build our sample of FR0 sources we took at first the SDSS/NVSS/FIRST
sample of radio galaxies by Best \& Heckman (2012) \footnote{Best \& Heckman (2012) built a sample of 18286 AGN by cross-correlating the seventh data release of the Sloan Digital Sky Survey (SDSS) with the NRAO VLA Sky Survey (NVSS) and the Faint Images of the Radio Sky at Twenty Centimeters survey (FIRST). The sample is selected at a flux density level of 5~mJy.}  and we applied the criteria listed below, following the approach of B15. This guarantees that we are considering LEG compact sources:

\begin{itemize}
\item [-] redshift z $\leq$0.15;
\item [-] compact in the FIRST images, corresponding to a radio size $\lesssim$10~kpc;
\item [-] FIRST flux $>$30~mJy (to ensures a higher fraction of X-ray detected objects); 
\item [-] LEG optical classification.
\end{itemize}

We obtained a list of 73 objects from which we excluded the four sources
classified as low-luminosity BL Lacs (Capetti \& Raiteri 2015).
We performed a search for X-ray observations of the remaining 69 sources
available in the public archives of the X-ray satellites currently
on-flight \footnote{\ttfamily{http://heasarc.gsfc.nasa.gov/cgi-bin/W3Browse/w3browse.pl}}
and found 15 objects. Some sources are the target of the X-ray pointing, some
others were serendipitous sources in the field of other targets. In order to
enlarge the sample, we also included two FR0s already presented in B15 
having public X-ray observations. These sources were not included in our starting sample
since they have FIRST fluxes $\sim$10~mJy.
Finally, during the
writing of the paper Tol1326-379, the first FR0 detected in $\gamma$-rays by
the \emph{Fermi} satellite (Grandi et al. 2016), was observed by \emph{Swift}
as a ToO and therefore it is considered in this work.
The entire sample of FR0s studied here is reported in Table~1.

Figure~1 shows the location of our 19 FR0s in the FIRST versus [OIII] diagram
adapted from B15. FR0s and 3CR/FRIs share the same range in L$_{\rm [OIII]}$, but FR0s 
have lower radio luminosities: this strong deficit in total radio emission places FR0s to the left of 3CR/FRIs (B15),
confirming that our selection criteria are valid.
Even considering low-luminosity radio galaxies such as FRICAT sources (Capetti et al. 2017), FR0s still occupy the left side
of the plot (see Figure~6 of Baldi et al. 2018a) forming a continuous distribution from FR0 to sFRICAT, FRICAT and 3CR/FRI sources. The low-luminosity BL Lacs of Capetti \& Raiteri (2015; see also Baldi et al. 2018b) are also shown in Figure~1 and have generally 1.4~GHz
radio luminosities higher than FR0s (see Section 4.2).  In the same plot also CoreG galaxies\footnote{CoreG galaxies are low-luminosity radio sources hosted by early-type galaxies and defined "core" on the basis of  the presence of a shallow core in their host surface brightness profile.} are reported (for more details see Balmaverde \& Capetti 2006; Baldi \& Capetti 2009).

The radio properties of our sample meet the FR0 criteria discussed by B15. 
The sources generally show flat radio spectra and are compact.
Indeed, as shown in Table~2, the ratios between the FIRST and NVSS
fluxes at 1.4~GHz are around 1, indicating that the extended component in these sources is
negligible. The core dominance {\it R} \footnote{{\it R} is defined
as the the ratio between 8.5~GHz (CLASSSCAT: Myers et al. 2003; Browne et al. 2003) and 1.4~GHz (NVSS) flux
densities.} is on average $\sim$30 times higher than 3CR/FRIs and overlaps with the FR0 values of B15 (Figure~2).
The paucity of information at radio frequencies higher than 1.4~GHz for the FRICAT sources
prevent us from comparing their core dominance with our X-ray sample of FR0s.
Finally, the radio spectral indices measured between 8.5~GHz (4.9~GHz) and
1.4~GHz are generally flat with a median value $\alpha_{r}$=-0.04 (see Table~2 for more details).

\begin{center}
\begin{table}
\caption{Radio parameters: (1) source name, (2) flux ratio between FIRST and NVSS fluxes, (3) radio spectral index (S$_{\nu} \propto \nu^{\alpha}$) between 8.5~GHz (The Cosmic Lens All Sky Survey, CLASSSCAT, Myers et al. 2003, Browne et al. 2003) and 1.4~GHz (NVSS) otherwise specified in the notes, (4) core dominance, {\it R}, defined as the ratio between 8.5~GHz (CLASSSCAT) and 1.4~GHz (NVSS) flux densities.}
\label{tab2}      
\begin{tabular}{l c r r}     
\hline\hline       
Source name               &F$_{\rm FIRST}$/F$_{\rm NVSS}$ &$\alpha_{r}$ &Log\it{R}\\
\hline
J004150.47-09      &0.74  &-0.13$^{a}$   &--     \\
J010101.12-00      &0.70  &-0.45$^{b}$   &-0.47$^{b}$ \\
J011515.78+00      &1.05  &-0.04$^{b}$   &-0.06$^{b}$  \\
J015127.10-08      &0.88  &--      &--    \\
J080624.94+17   	  &0.96  &--      &--    \\
J092405.30+14      &0.95  &--      &--    \\
J093346.08+10      &0.80  &-0.47   &-0.37 \\
J094319.15+36      &0.99  &0.82   &0.65  \\					
J104028.37+09      &0.94  &-0.63$^{a}$    &--   \\                        
J114232.84+26   	  &0.94  &0.11    &0.09   \\
J115954.66+30    	  &1.06  &-0.01   &-0.01 \\
J122206.54+13      &0.99  &0.28     &0.22  \\
J125431.43+26    	  &0.98  &-0.47   &-0.37  \\
Tol1326-379        & --   &0.37$^{c}$    &-0.42 \\
J135908.74+28    	  &1.12  &-0.12    &-0.09  \\
J153901.66+35      &0.92  &-0.10   &-0.08 \\				      
J160426.51+17      &0.75  &0.15    &0.12  \\
J171522.97+57      &0.86  &-0.43   &-0.34 \\
J235744.10-00      &0.65  &-0.67$^{b}$   &-0.58$^{b}$ \\

\hline\\
\multicolumn{4}{l}{$^{a}$ $\alpha_{r}$ between 4.9~GHz (JVASPOL or PMN) and 1.4~GHz.}\\
\multicolumn{4}{l}{$^{b}$ Values of $\alpha_{r}$ and {\it R} are from B15.}\\
\multicolumn{4}{l}{$^{c}$ The value of $\alpha_{r}$ is from Grandi et al. (2016).}\\
\end{tabular}
\end{table}
\end{center}


\begin{figure}
\centering
\includegraphics[width=\columnwidth]{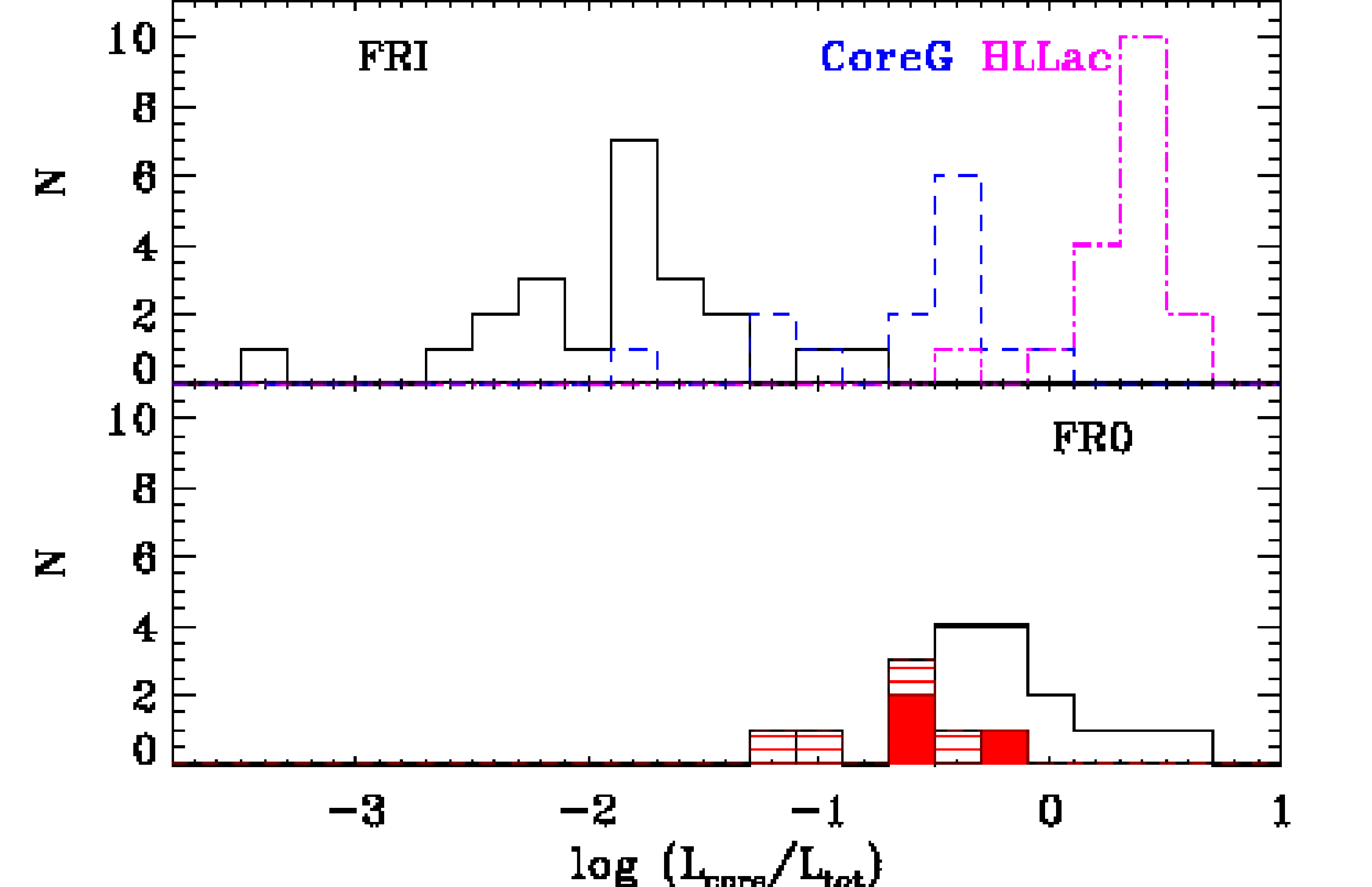}
\caption{Histograms of the core dominance (adapted from B15) for: {\it upper panel}- 3CR/FRIs
  ({\it black solid line}), CoreG ({\it blue dashed line}), low-luminosity BL
  Lacs ({\it magenta dot-dashed line}). {\it Lower panel}- FR0s of B15 ({\it
    red ticked line}) and our sample of FR0s ({\it black solid line}). The
  {\it red filled} histograms are the sources in common with B15.}
\label{fig1b}
\end{figure}

\section{X-ray observations and analysis}

\subsection{Data reduction}
Data were collected from different X-ray satellites.
In particular, 7 sources
were observed with \emph{Chandra}, 4 with \emph{XMM-Newton}
and 8 with \emph{Swift}/XRT. The observation log is in Table~1.
Several FR0 sources are not the primary target of the observation but are in
the field of view of the pointing. The offset, i.e. the distance from the
center of the source cone \footnote{See also
  \ttfamily{https://heasarc.gsfc.nasa.gov/W3Browse/w3browse-help.html$\#$distance$\_$from$\_$center}}
is also reported in Table~1. When more than one observation was available, we
chose the one with the smaller offset or the longer exposure.  We reduced data
for all the sources but one, i.e. J235744.10-00, since it is part of the 3XMM-DR6
catalog (Rosen et al. 2016).

All \emph{Chandra} observations were performed using CCD, both ACIS-S and ACIS-I.  Data were reprocessed using CIAO version 4.7 with calibration database CALDB version 4.6.9 and applying standard procedures. Table~A1 reports the extraction regions chosen for the nuclear and background spectra of each source.
Data were then grouped to a minimum of 15 counts per bin over the energy range 0.5-7~keV. None of the seven sources pointed by \emph{Chandra} is affected by pile-up.

For the \emph{XMM-Newton} observations we reduced and analyzed data from the EPIC-pn camera using SAS version 14.0 and the latest calibration files. Periods of high particle background were screened by computing light curves above 10~keV. Extraction regions for the source and background spectra are reported in Table~A1.
Data were then grouped to a minimum of 15 counts per bin over the energy range 0.5-10~keV for two out of the three sources. For the source J235744.10-00 the 0.2-12~keV flux was directly taken from the 3XMM-DR6 catalog, and extrapolated to 2-10~keV 
assuming the same spectral slope adopted in the catalog.

\emph{Swift}/XRT data were reduced using the online data analysis tool provided by the ASI Space Science Data Center (SSDC) \footnote{\ttfamily http://swift.asdc.asi.it/}. The only exception is Tol1326-379 that was observed as a ToO during the writing of the paper and the data were processed and analyzed using standard XRT tools (xrtpipeline v.0.13.2  and caldb v.1.0.2). Source spectra for each observation were extracted from a circular region of 20$''$ radius, while the background was taken from an annulus with an inner radius of 40$''$ and outer radius of 80$''$. Spectra were grouped to a minimum of 5 (or 3) counts per bin in the energy range 0.5-10~keV. In the case of the source J153901.66+35 no grouping was applied.

\subsection{Imaging analysis}

The inspection of the X-ray images indicates that 6 sources
are in a dense environment (Figure~3). Four lie at the outskirts of a cluster
of galaxies (i.e. J160426.51+17, J092405.30+14, J135908.74+28, J011515.78+00), J004150.47-09 is located at the
centre of the cluster Abell85 and J171522.97+57 is the brightest member in a compact
group of more than 13 galaxies (Pandge et al. 2012).  

For the sources for which a clear extension cannot be confirmed by the X-ray
images, information on the environment was checked in the literature (see
Table~3). Other 4 sources were found in clusters or compact groups (CG), i.e.,
J093346.08+10, J122206.54+13, J080624.94+17, J115954.66+30 (Diaz-Gimenez et al. 2012; Owen et al. 1995;
Koester et al. 2007). In summary, we found that at least 50$\%$ of the FR0s of our sample
is in a dense environment. This value should be considered as a lower limit to the fraction 
of FR0s in dense environments in our sample since the analysis suffers from a bias
due to the nature of the sample mainly consisting of X-ray serendipitous
sources. However, we are aware that strong conclusions on the environment of FR0 as a class
cannot be drawn with the available data.

\begin{figure*}
\centering
\includegraphics[width=6cm,height=6cm]{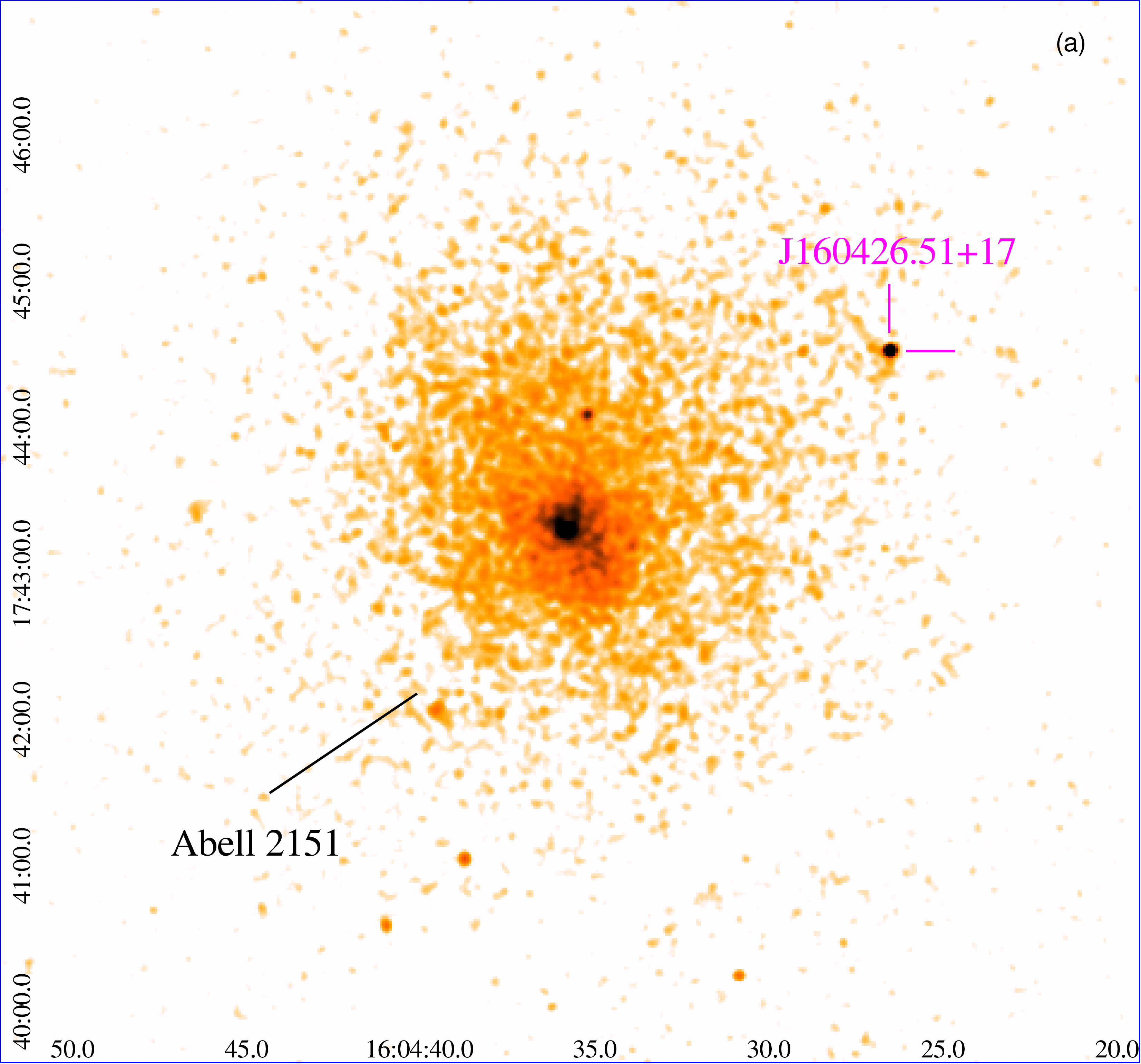}
\hspace{0.1cm}
\includegraphics[width=6cm,height=6cm]{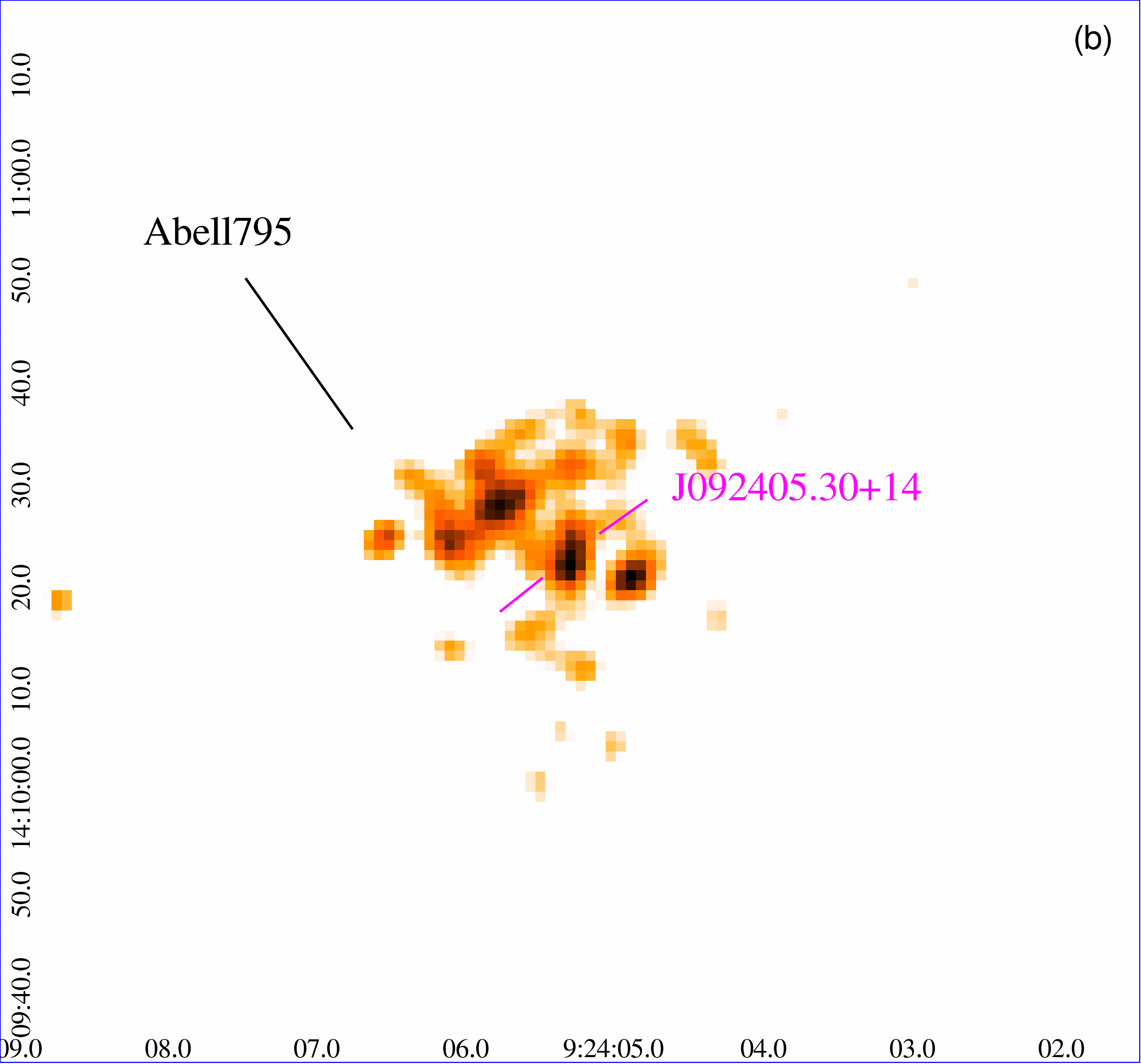}
\hspace{0.1cm}
\includegraphics[width=6cm,height=6cm]{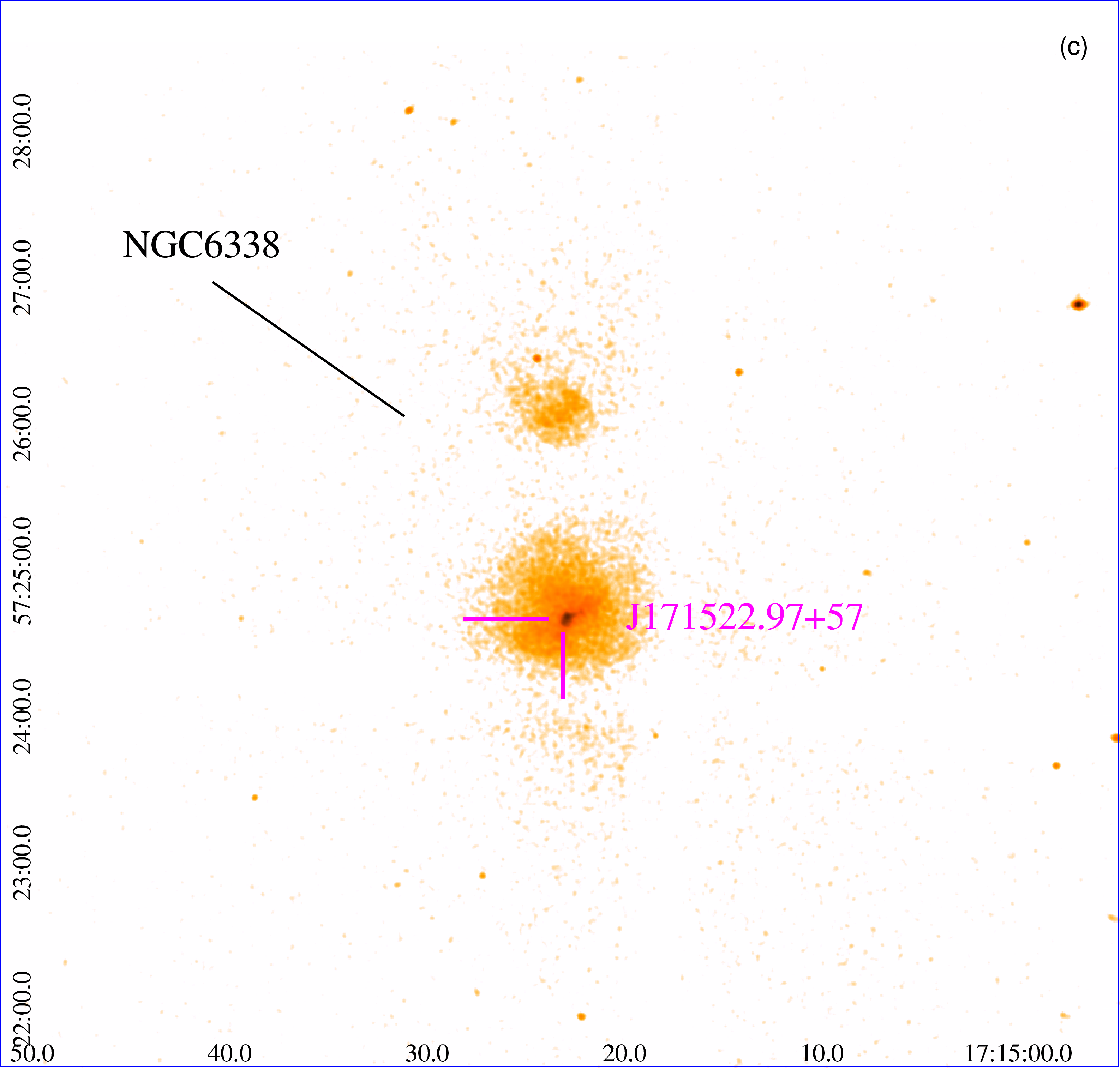}
\hspace{0.1cm}
\includegraphics[width=6cm,height=6cm]{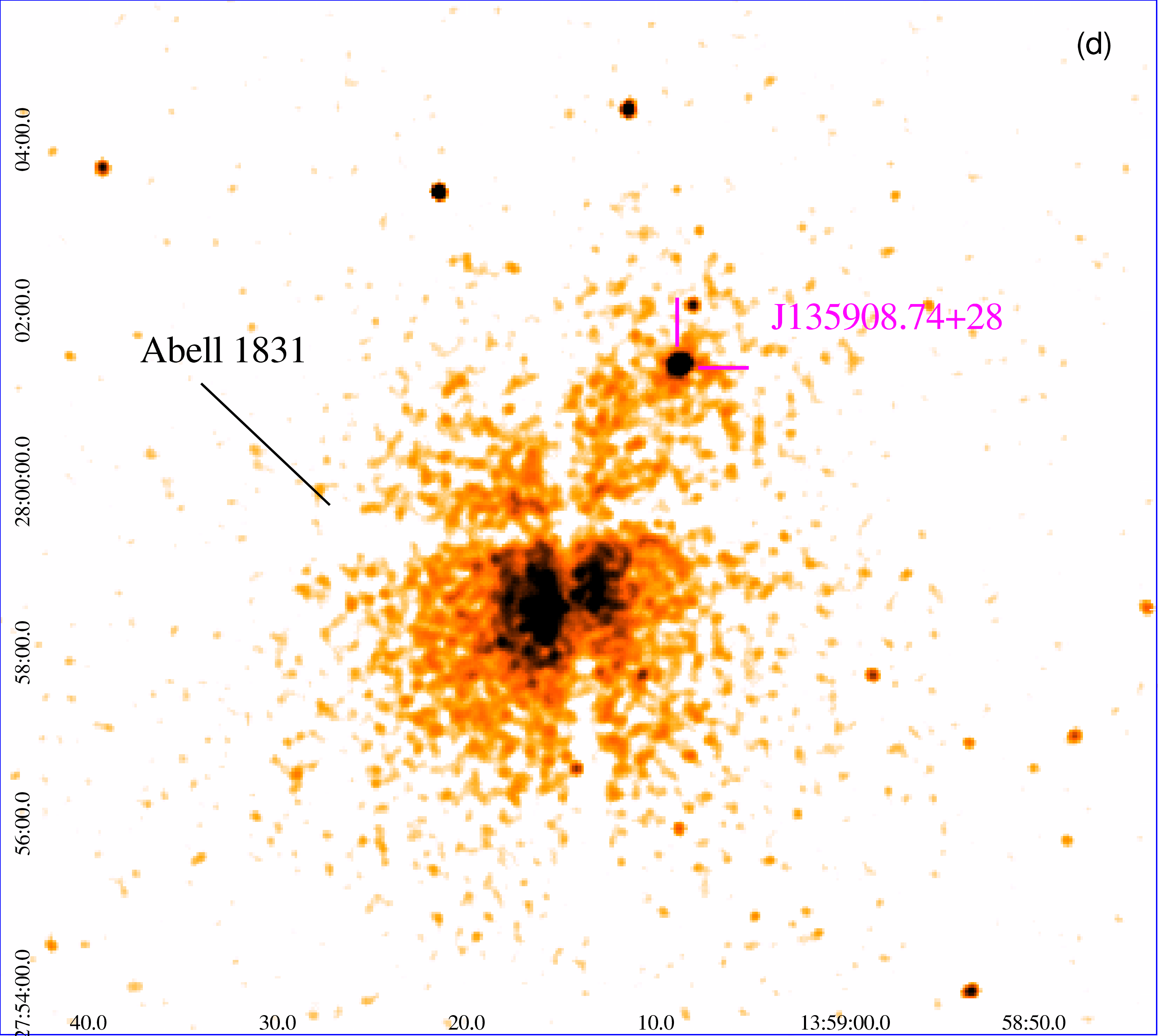}
\hspace{0.1cm}
\includegraphics[width=6cm,height=6cm]{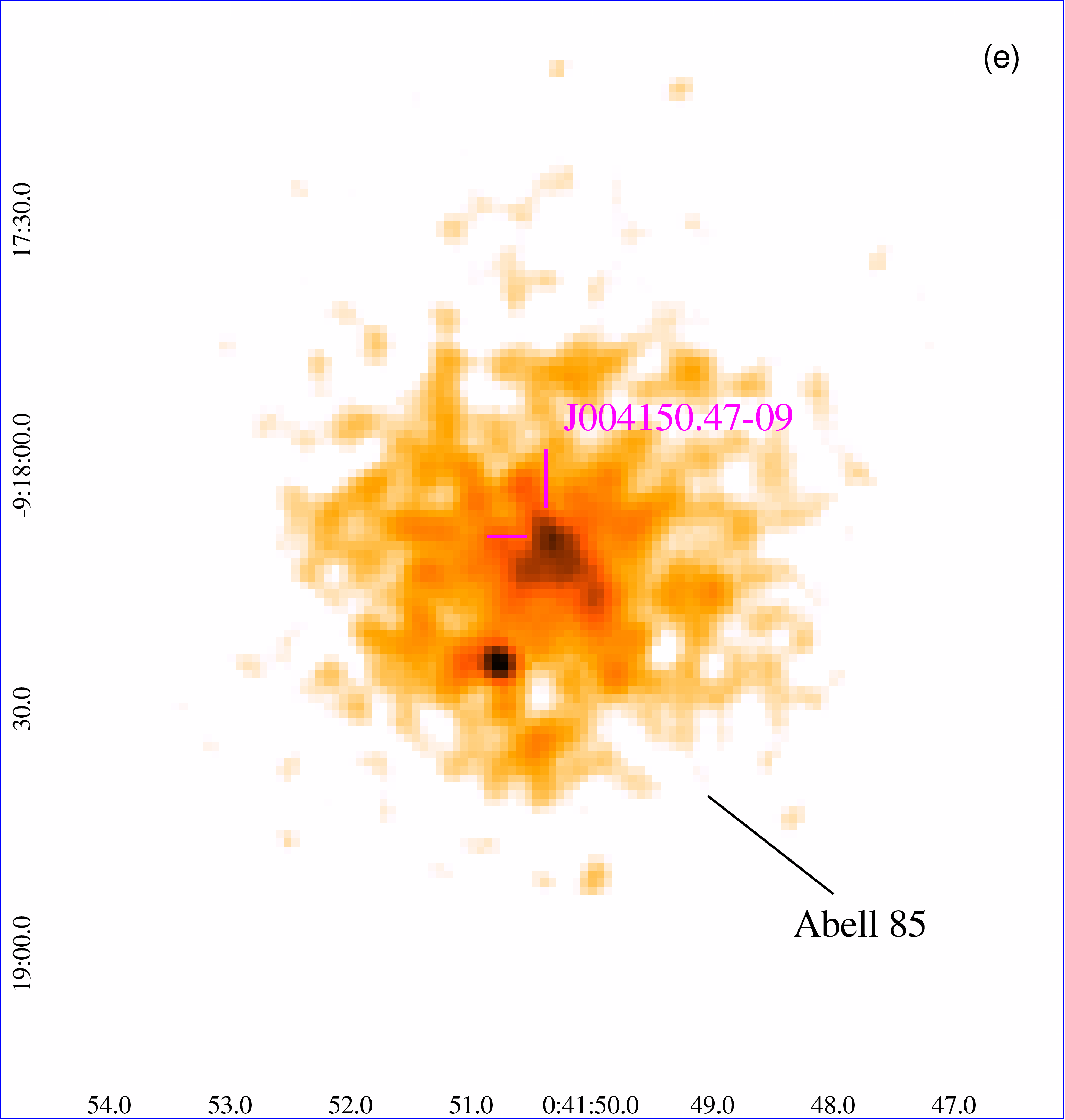}
\hspace{0.1cm}
\includegraphics[width=6cm,height=6cm]{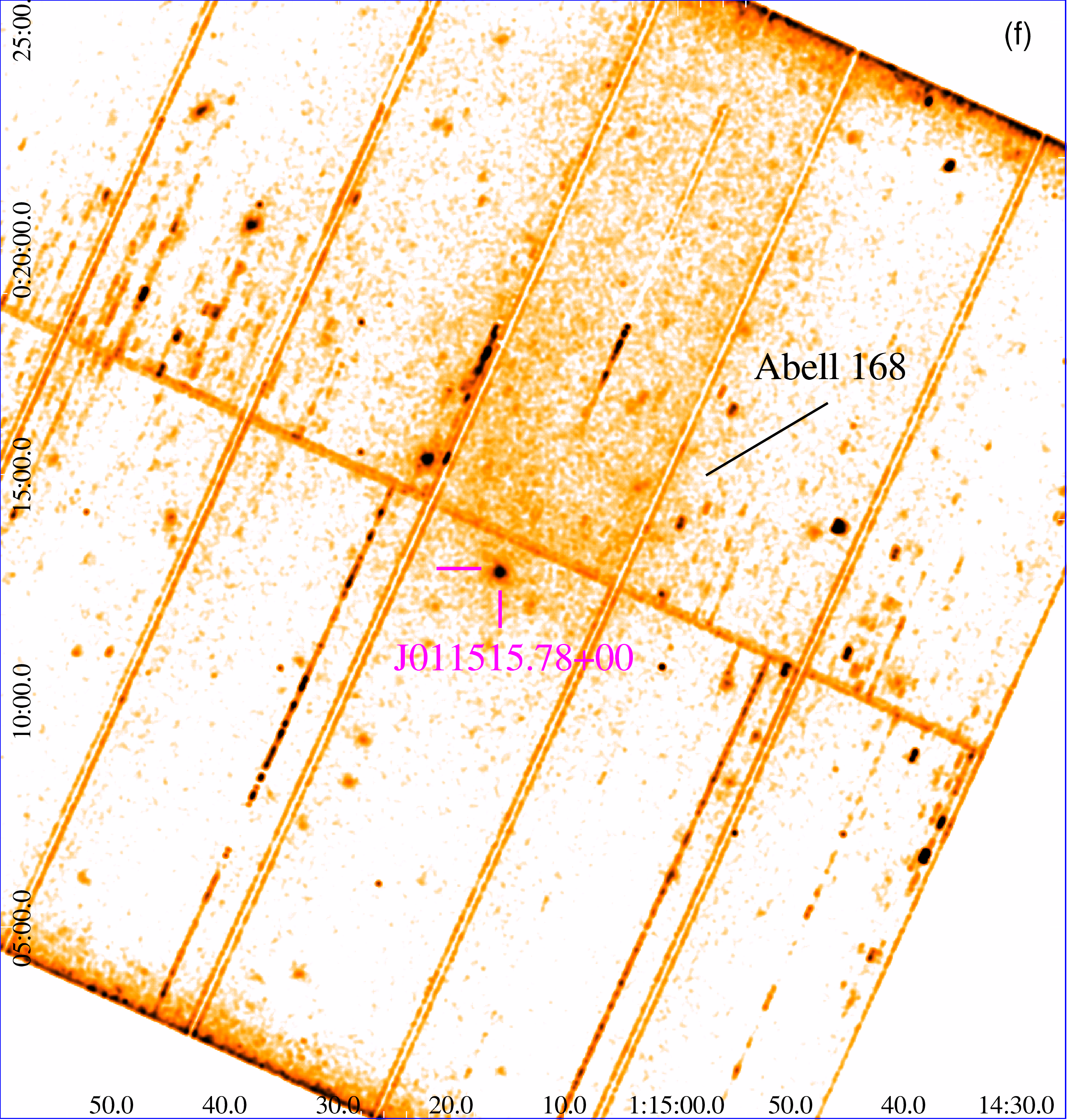}
\caption{Images of the clusters found in the X-ray band. \emph{Panels
    (a)-(d)}: \emph{Chandra} 0.3-7~keV images of the sources J160426.51+17, J092405.30+14,
  J171522.97+57, J135908.74+28 . The FR0 is labelled in magenta while the name of the cluster is in black. \emph{Panel (e)}: \emph{Chandra} 3-7~keV image of the source J004150.47-09. \emph{Panel (f)}: \emph{XMM-Newton}/pn image of the
  source J011515.78+00. All images have been smoothed with a Gaussian with kernel
  radius 3.}
\label{fig}
\end{figure*}

\subsection{Spectral analysis}
The spectral analysis was performed using the {\small XSPEC} version 12.9.0
package. We applied a $\chi^{2}$ statistics to spectra binned to a minimum of
at least 15 counts per bin. When the grouping was smaller a C-statistics was
adopted. Errors are quoted at 90$\%$ confidence for one interesting parameter
($\Delta\chi^{2}$=2.71).  The summary
of the best-fit spectral results is reported in Table~3. Notes on single
sources can be found in Appendix~A.

Spectral fitting was performed in the energy range 0.5-7~keV (\emph{Chandra})
and 0.5-10~keV (\emph{XMM-Newton} and \emph{Swift}). The X-ray luminosities 
presented throughout the paper are calculated in the 2-10~keV range in order to make
a direct comparison with the literature.
As a baseline model, we considered a power-law convolved with the Galactic
column density (Kalberla et al. 2005). In 4 out of 6 sources for which we
could directly observe the cluster in the X-ray images (Figure~3), residuals
showed evidence for the presence of a soft component. Therefore we included a thermal
model ({\small APEC}). 
A thermal component was also required in other three FR0s. The presence of a compact group was attested in J093346.08+10 and J115954.66+30 checking the literature. For the third one (i.e J015127.10-08) no information on the environment was found.
The nature of the soft X-ray emission is however uncertain. It could be due to an extended intergalactic medium  (that can not be revealed because of poor X-ray spatial resolution and/or short exposure time) or 
related to the hot corona typical of early-type galaxies (Fabbiano, Kim \& Trinchieri 1992).

We could measure the power-law photon indices $\Gamma$ for 7 out of 18
objects. The spectral slopes are generally steep, with a mean value
$<\Gamma>$=1.9 and a standard deviation of 0.3.  
When it was not possible to leave the photon index free, it was fixed to a value of 2. We
checked whether different values of the photon index lead to significant
changes in the estimate of the fluxes. We found that for $\Gamma$ ranging
between 1.5 and 2.5 the fluxes are consistent within the errors.  
In four cases the low statistics did not allow us to constrain the
power-law component and to exclude the presence of thermal emission, therefore we assumed
a simple power-law ($\Gamma$=2 fixed) as the best-fit model and we adopted the
resulting 2-10~keV flux as upper limit for the nuclear component.

Generally, the X-ray spectra of our sample do not show evidence for intrinsic 
absorption. Indeed, the addition of an intrinsic absorber component does not improve
significantly the fit. An upper limit to this component can be estimated only for three out of 18 sources (see Appendix~A). 
Therefore, we tend to favor the scenario in which the circum-nuclear environment of FR0
is depleted of cold matter, similarly to FRIs (Balmaverde et al. 2006; Baldi \& Capetti 2008, 2010; Hardcastle et al. 2009). 

The analyzed FR0s have X-ray nuclear luminosities covering three orders of magnitude
10$^{40-43}$~erg~s$^{-1}$. The  average value including the upper limits is $<$LogL$_{\rm X}>$=41.30 (see Figure~4 {\it upper panel}).

\section{Discussion}

\subsection{Compact versus extended low-excitation radio galaxies}

FR0/LEGs and FRI/LEGs reside in similar galaxies and share similar nuclear optical properties 
(B15 and references therein). Given that low-ionization optical spectra can be also produced 
by shocks or old stellar population emission (Binette et al. 1994; Sarzi et al. 2010; Capetti \& Baldi 2011; Balmaverde \& Capetti 2015; Mingo et al. 2016),
this X-ray study is a key tool to compare FR0 and FRI properties taking advantage
of an energy band directly
related to the nuclear emission processes.  

We compared the X-ray luminosities of our sample to those of 35 FRI radio
galaxies belonging to the 3CR/3CRR catalogs and having X-ray data available.
The 2-10~keV luminosities of FRIs are from literature or obtained by a direct
analysis of the data stored in the public archives (see Appendix~B for
details).
Various samples of FRIs can be considered for the comparison with 
compact radio sources. For example the 3C sample includes $\sim$30 FRIs,
the B2 sample is formed by $\sim$100 radio galaxies, about half of them being
FRIs. The recent FRICAT catalog is formed by 219 sources, selected from the SDSS/NVSS surveys,
with a flux limit of 5~mJy at 1.4~GHz.
However, the multi-wavelength information is rather limited and, in
particular, the coverage of X-ray observations is very small for all samples
except for the 3C for which {\it Chandra} data are available for all sources up to
z=1 (Massaro et al. 2010, 2012, 2013). Being selected with a rather high flux threshold (9~Jy at
178~MHz), they represent the tip of the iceberg of the FRI population. While
this means that the view of FRIs offered by the 3C sample is limited, they
certainly provide us with a benchmark against which compare the properties
of the compact radio sources.

The two distributions in Figure~4 ({\it upper panel}) clearly overlap: the two sample tests univariate program in the ASURV package
(TWOST, Feigelson \& Nelson 1985; Isobe, Feigelson \&
Nelson 1986) applied to the data (including upper limits) confirms their similarity (P$_{\rm ASURV}$=0.76)\footnote{Probability value according
to the Gehan's generalized Wilcoxon test.}. We assume that 
P=0.05 is the probability threshold to rule out the hypothesis that the two samples are
drawn from the same parent population.
This result indicates a strong correspondence between the X-ray cores 
of low-excitation FR0 and FRI radio galaxies\footnote{A similar result is obtained even excluding upper limits and applying a 
Kolmogorov-Smirnov test to the data (P$_{\rm KS}$=0.45).}.

This point is strengthened by Figure~4 ({\it lower panel}) where the X-ray
(2-10~keV) and radio core (5~GHz) luminosities of FR0s and FRIs are plotted
together. Apart from the two objects having measurements at 4.9~GHz (see Table~2), the luminosities at 5~GHz of the compact sources were extrapolated from
1.4~GHz (FIRST) data considering the radio spectral slopes reported in
Table~2.  The core luminosities of the extended radio galaxies are from
literature (Buttiglione et al. 2010, 2011; Hardcastle et al. 2009). The two
samples occupy the same area in the plot: the generalized Kendall's $\tau$
test (ASURV package: Isobe et al. 1986) gives a probability of correlation
greater than $99.99\%$ and $99.95\%$ for FRIs and FR0s (including upper
limits), respectively.

We also tested the possible influence of redshift in driving this correlation
estimating a partial rank coefficient \footnote{The partial rank coefficient
  estimates the correlation coefficient between two variables after removing
  the effect of a third. If A and B are both related to the variable z, the
  partial Kendall's $\tau$ correlation coefficient is:
  $\tau_{AB,z}=\frac{\tau_{AB}-\tau_{Az}\tau_{Bz}}{\sqrt{(1-\tau_{Az}^{2})(1-\tau_{Bz}^{2})}}$.}. The
effect is negligible and the value of the correlation coefficient does not
change significantly.

The correlation between L$_{\rm X}$--L$_{\rm 5GHz}$ already found for 3CR/FRIs (Balmaverde
et al. 2006), it is now attested for FR0s pointing towards a jet origin of
both the radio and X-ray photons. As reported in Section 3.3, an intrinsic absorber is not
required by the fit, at least for those sources having good quality X-ray spectra. Therefore,
it is unlikely that the jet-related X-ray emission observed in FR0s is the unabsorbed component
of a HEG-like spectrum. A similar result was obtained by  Mingo et al. (2014), who concluded that the LEGs of their 2~Jy sample 
cannot be interpreted as simple heavily obscured HEGs (see also Baldi et al. 2010). This point is further strengthened below by 
the estimate of the Eddington-scaled luminosities for our FR0 sample.

From the stellar velocity dispersion relation of Tremaine et
al. (2002)\footnote{log(M$_{\rm BH}$/M$_{\odot}$)=(8.13$\pm$0.06)+(4.02$\pm$0.32)log($\sigma$/200~km~s$^{-1}$)} we estimated the black hole masses (M$_{\rm BH}$) of our sample of FR0s.  The values of M$_{\rm BH}$ range between $\sim 10^{8}$ and $\sim
10^{9}$~M$_{\odot}$ (see Table~4).  From the relation
L$_{\rm bol}$=3500~L$_{\rm [OIII]}$ (Heckman et al. 2004) we derived the bolometric
luminosities and successively the Eddington-scaled luminosities
($\dot{L}$=L$_{\rm bol}$/L$_{\rm Edd}$) given in Table~4. These estimates for
the FR0s of our sample correspond to low values of $\dot{L}$ ($\sim
10^{-3}-10^{-5}$) typical of inefficient accretion modes (ADAF-like, Narayan
\& Yi 1994, 1995) and similar to those found for FRIs (see also Table~B1) and for 2~Jy LEGs (Mingo et al. 2014).

This result strengthens
the interpretation of a non-thermal origin of the high-energy nuclear emission
in low-excitation compact sources, already suggested by the correlation between radio and X-ray
emissions.

\subsection{Compact radio galaxies versus BL Lac objects}

The radio compactness of FR0s is due to the lack of extended emission and it is
not related to a Doppler boosting of the jet radiation. This is evident in the
compact radio galaxies that show emission lines with large equivalent widths in
their optical spectra (such as Tol1326-379, Grandi et al. 2016) but it is less
straightforward in objects overwhelmed by the galaxy emission. Indeed, in this
case, the stellar population dominates the emission hiding the nature of the
underlying AGN that could be both a low-luminosity BL Lac with an extended jet
pointed towards the observer, or a genuine compact radio galaxy.

In Figure~1 it is evident that FR0s and low-luminosity BL Lacs occupy different
regions of the L$_{\rm [OIII]}$-L$_{\rm 1.4~GHz}$ plane. The two classes have similar
emission-line luminosities but BL Lacs are more powerful in the radio band.
This is expected since L$_{\rm [OIII]}$ is an isotropic indicator of the AGN
luminosity, while the radio emission suffers from relativistic effects.

The ratio between the [OIII]$\lambda$5007 line and the 2-10~keV luminosities
(R$_{\rm [OIII]}$=L$_{\rm [OIII]}$/L$_{\rm (2-10~keV)}$) can be considered a useful tool to
distinguish misaligned and aligned jets.  We then collected [OIII]
luminosities of FRIs and FR0s from literature (Buttiglione et al. 2010, 2011;
Hardcastle et al. 2009; Leipski et al. 2009) and from SDSS/DR7
survey \footnote{\ttfamily{http://classic.sdss.org/dr7/}}, respectively (see
Appendix~B and Table~4). For the [OIII]-line luminosity of BL Lacs we refer to
the work of Capetti \& Raiteri (2015), while the X-ray 2-10~keV luminosity was
directly obtained from the Swift/XRT instrument using the SSDC online data
analysis tool.

The R$_{\rm [OIII]}$ average values (including upper limits \footnote{We used the
  Kaplan-Meier (KM) estimator in ASURV to derive the average values of
  R$_{\rm [OIII]}$ in the presence of censored data.}) for FR0s and FRIs are
consistent ($<$R$_{\rm [OIII], FR0}$$>$=-1.6$\pm$0.2 and $<$R$_{\rm [OIII],
  FRI}$$>$=-1.7$\pm$0.2). This is statistically attested by the TWOST test in
ASURV (see Section 4.1 for details) that turns out with a probability
P$_{\rm ASURV}$=0.9.  On the contrary, the Doppler boosting of the X-ray emission
shifts the BL Lacs to lower values $<$R$_{\rm [OIII], BL Lacs}$$>$=-3.3$\pm$0.2.

\subsection{Compact radio galaxies versus young sources}

The comparison between FR0s and young radio sources in the X-ray band suffers from the paucity of dedicated studies in this field (Kunert-Bajraszewska et al. 2014 and references therein). The samples of young sources for which high-energy information are available include mainly powerful GPS and CSS. These sources are generally different from ours being characterized by high X-ray (2-10~keV), radio (5~GHz) and [OIII]-line luminosities typical of AGN with efficient accretion rates (Guainazzi et al. 2006; Vink et al. 2006; Labiano 2008; Siemiginowska et al.2008; Tengstrand et al. 2009).
Moreover, 16 CSO sources recently studied by  Siemiginowska et al. (2016) in X-rays showed  spectra generally flat and absorbed by intrinsic column densities (see also Ostorero et al. 2017). 

Our sources seem more similar to three low-luminosity compact sources (LLC;
Kunert-Bajraszewska \& Thomasson 2009) discussed by Kunert-Bajraszewska et
al. (2014) and classified as LEG. Their radio and X-ray luminosities (see
Table~2 of their work) locate these sources in our correlation strip shown in
Figure~4 ({\it lower panel}). The authors suggest that such LLC are
intermittent radio sources rather than young objects evolving in FRIs, in line
with the recent demographic study of Baldi et al. (2018a). They showed that the
space density of FR0s in the local Universe (z$<$0.05) is larger by a factor
of $\sim$5 than FRIs, definitively rejecting the hypothesis that FR0s
are young radio galaxies that will all eventually evolve
into extended FRI radio sources.

\begin{figure}
\centering
\includegraphics[width=8cm, height=8cm]{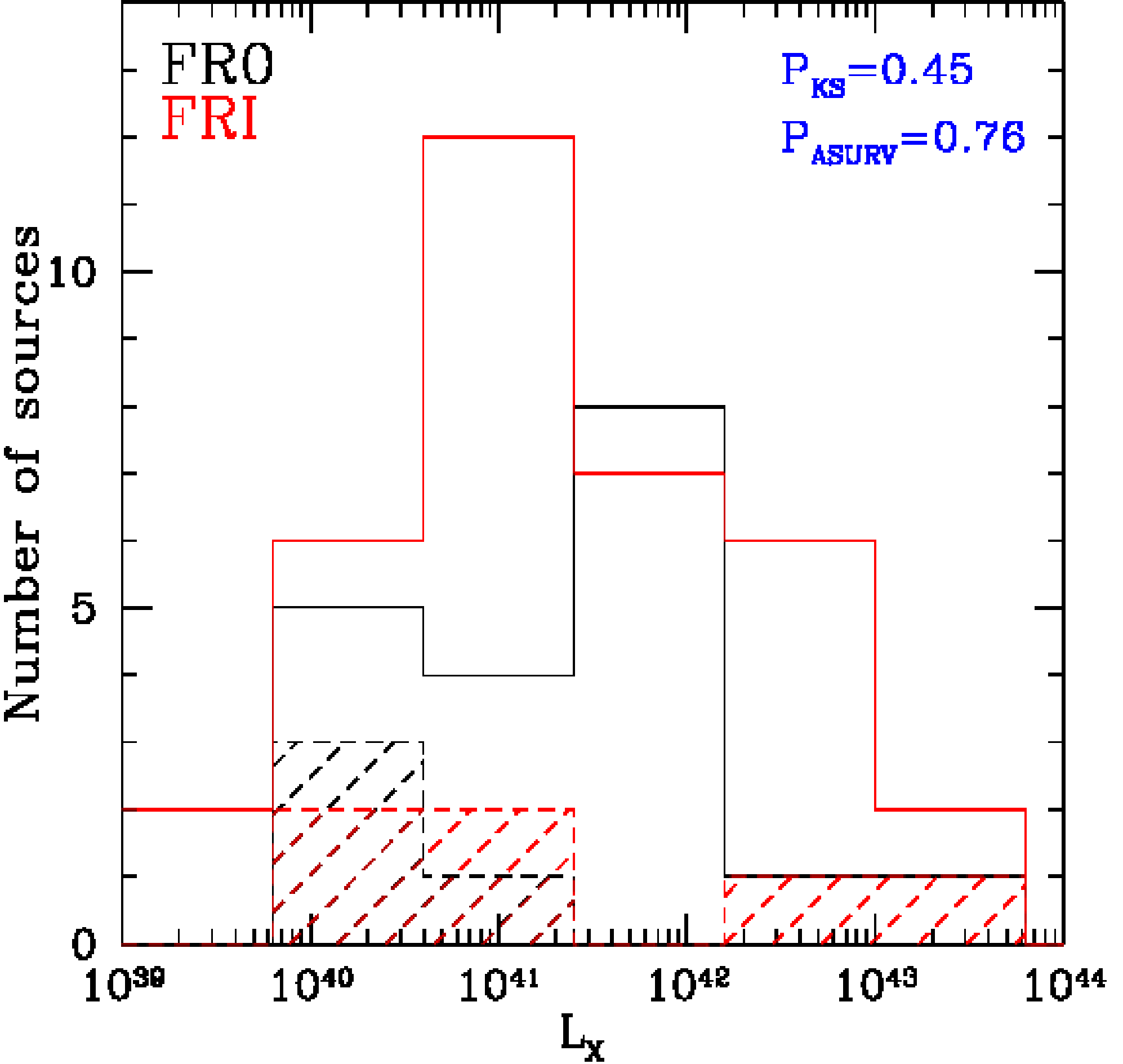}
\includegraphics[width=8.5cm, height=8.5cm]{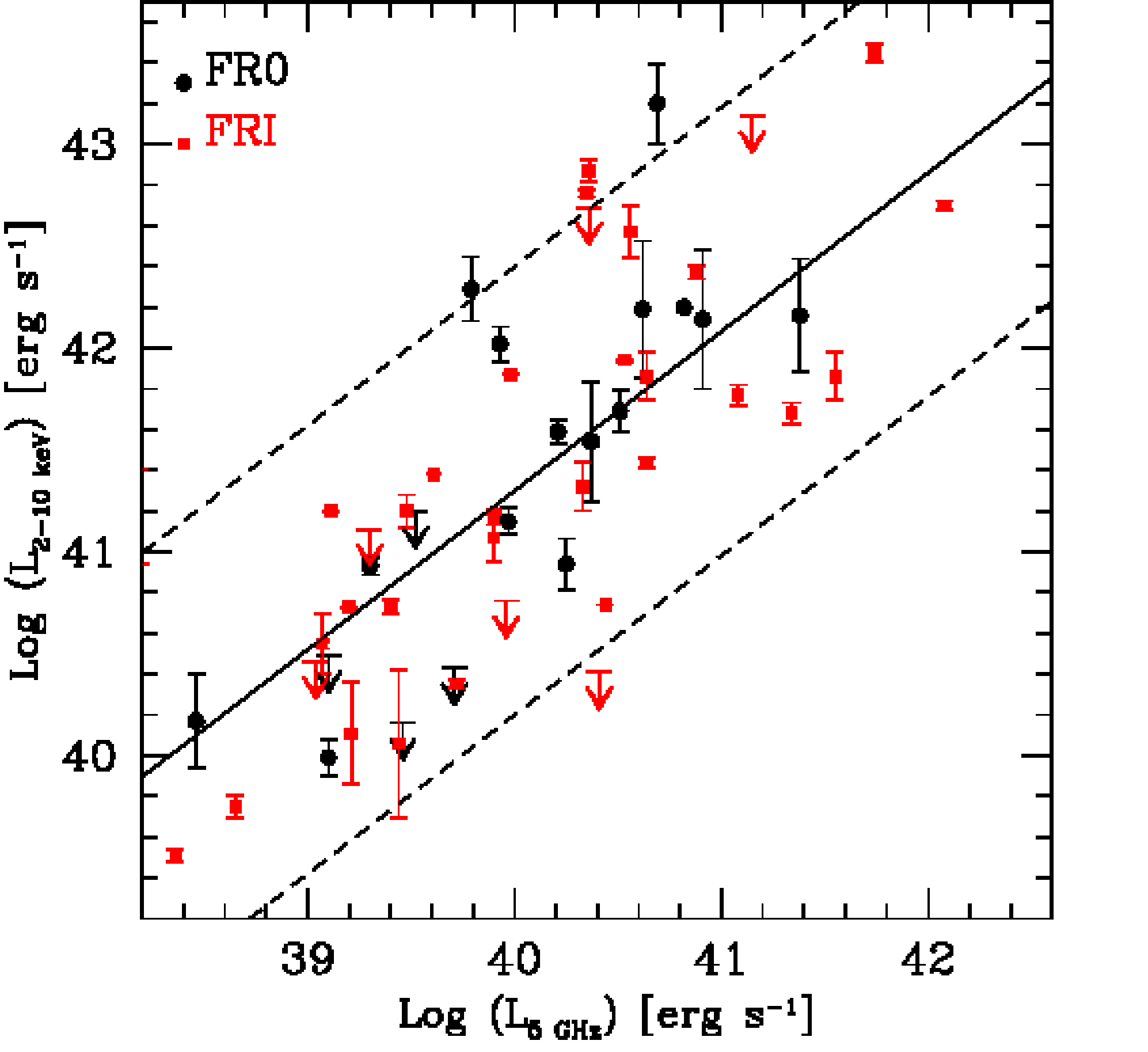}
\caption{{\bf Upper panel}: histogram of the  X-ray luminosity for the FR0s of our sample ({\it black solid line}) and the comparative sample of FRIs ({\it red solid line}). No significant difference is observed between the two samples. The {\it dashed histograms} represent upper limits in each sample. In the plot the probabilities from the Kolmogorov-Smirnov test (P$_{\rm KS}$) and the TWOST test in ASURV (P$_{\rm ASURV}$) are also reported. {\bf Lower panel}: 2-10~keV X-ray luminosity versus 5~GHz radio core luminosity for FR0s ({\it black circles}) and FRIs ({\it red squares}). Arrows indicate upper limits. 
The {\it black solid line} is the linear regression for the overall sample of  FR0s and FRIs (excluding the upper limits): LogL$_{\rm X}$=(7.8$\pm$0.6)+(0.8$\pm$0.1)LogL$_{\rm 5GHz}$ (see also Hardcastle et al. 2009). The {\it black dashed lines} are the uncertainties on the slope.}
\label{fig4}
\end{figure}

\section{Summary and conclusions}

 We analyzed 19 FR0s  selected according to the criteria discussed in Section~2 and having public X-ray observations. Most of
the sources have short exposures and/or large offsets. In spite of the limited
quality of the data, our analysis allowed us to characterize for the first
time FR0 sources at high-energies.

FR0s have X-ray luminosities (2-10~keV) between 10$^{40}$-10$^{43}$
~erg~s$^{-1}$, comparable to FRIs.  The clear correlation between radio
and X-ray luminosities observed in both compact and extended objects, favours
the interpretation of a non-thermal origin of the 2-10~keV photons. In
agreement with FRIs, the high-energy emission in FR0s is produced by the jet.
Moreover, the high black
hole masses (10$^{8}$-10$^{9}$~M$_{\odot}$) and the small values of the
bolometric luminosities, as deduced from the [OIII]-emission line, suggest an
inefficient accretion process (ADAF-like) at work also in the compact sources.
These results confirm that the nuclear properties of FRIs and FR0s are similar 
and that the main difference between the two classes remains the lack of extended emission in FR0s.

We exclude important beaming effects in the X-ray spectra of FR0s on the basis
of the ratio between the [OIII]$\lambda$5007 line and the 2-10~keV luminosity
(R$_{\rm [OIII]}$).  While the [OIII]-line luminosity is expected to be emitted isotropically,
the X-ray radiation could be amplified by Doppler boosting effects. While FR0s
and FRIs have similar R$_{\rm [OIII]}$ ($\sim$-1.7), low-luminosity BL Lacs, whose X-ray radiation
is beamed, have smaller values (R$_{\rm [OIII]}$ $\sim$-3.3).
 
A comparison with the X-ray properties of young sources is limited by the
paucity of X-ray studies of GPS, CSS and CSO. Considering the available data,
we do not find spectral similarities between these sources and
FR0s. Generally, the studied young sources have higher X-ray luminosities
(they are probably associated with an efficient accretion disc) and often show
signatures of intrinsic absorption. Therefore, FR0s could be different sources
characterized by intermittent activity, as in the case of J004150.47-09 and/or by
jets with intrinsic properties that prevent the formation and evolution of extended
structures.


\begin{landscape}
\begin{table}
\caption{Main spectral parameters of the FR0 sample: (1) Name of the source, (2) redshift, (3) best-fit spectrum, (4) Galactic hydrogen column density, (5) power-law photon index, (6) temperature of the thermal component ({\small APEC}), (7) flux of the thermal component in the soft X-ray band (0.5-5~keV) corrected for Galactic absorption, (8)  flux of the non-thermal component in the 2-10~keV band corrected for Galactic absorption, (9) $\chi^{2}$/degrees of freedom reported when the grouping was $\geq$15. When the grouping was $<$15 the C-statistics was applied, (10) type of environment, (11) references for the environment.}
\label{tab3}      
\footnotesize
\begin{tabular}{l c c c c c c c c cc}     
\hline\hline       
Source name     &z          &Best spectrum     &N$_{H,Gal}$             &$\Gamma$   &kT   &F$_{th,0.5-5~keV}$     &F$_{nuc,2-10~keV}$        &$\chi^{2}$/d.o.f.   &Environment      &Reference$^{d}$\\ \\
              &            &               &(atoms~cm$^-2$)      & &(keV)   &(erg~cm~$^{-2}$~s$^{-1}$)  &(erg~cm~$^{-2}$~s$^{-1}$)   &        &   \\
\hline
J004150.47-09  &0.055   &PL              &2.8$\times$10$^{20}$  &2.0 (fix)          &-                                  &-          &(4.5$^{+2.9}_{-2.8}$)$\times$10$^{-14}$     &1.4$/$3 &Cluster      &1 \\
J010101.12-00  &0.097	 &PL	             &3.2$\times$10$^{20}$  &2.0 (fix)	                 &-            &-                  &<6.6$\times$10$^{-15}$                              &- &?            &-\\
J011515.78+00  &0.045  &APEC+PL  &3.0$\times$10$^{20}$  &2.0$\pm$0.2     &0.8$\pm$0.1    &(1.5$\pm$0.4)$\times$10$^{-14}$      &(2.9$^{+0.5}_{-0.4}$)$\times$10$^{-14}$    & 66/63    &Cluster      &1\\
J015127.10-08  &0.018  &APEC+PL  &2.3$\times$10$^{20}$  &2.0 (fix)    &0.2$^{+0.3}_{-0.1}$     &(1.4$^{+113}_{-1.3}$)$\times$10$^{-13}$    &$<$4.2$\times$10$^{-14}$   &-         &?            &-       \\
J080624.94+17  &0.104  &PL      &3.3$\times$10$^{20}$  &2.0 (fix)       &-           &-                             &(3.2$^{+1.7}_{-1.4}$)$\times$10$^{-13}$    &-     &Cluster      &2 \\
J092405.30+14  &0.135  &APEC+PL  &3.3$\times$10$^{20}$  &1.8$\pm$0.3   &1.3$^{+1.4}_{-0.6}$      &(5.9$^{+19}_{-5}$)$\times$10$^{-15}$    &(3.2$^{+1.1}_{-2.8}$)$\times$10$^{-14}$    & 3/9&Cluster   &1   \\
J093346.08+10  &0.011	  &APEC+PL  &3.1$\times$10$^{20}$  &2.0(fix)      &$<$0.35      &$<$1.2$\times$10$^{-13}$             &(6.2$^{+2.7}_{-2.1}$)$\times$10$^{-14}$.    &-	&CG$^{c}$     &3\\   
J094319.15+36  &0.022  &PL      &1.1$\times$10$^{20}$  &2.3$\pm$0.4          &-               &- 		         &(7.9$^{+2.3}_{-2.0}$)$\times$10$^{-14}$    &-  &?            &-  \\
J104028.37+09  &0.019  &PL     &2.6$\times$10$^{20}$   &2.2$\pm$0.4         &-                &-                 &(1.2$\pm$0.2)$\times$10$^{-14}$                & 3/7    &Isolated	   &4  \\
J114232.84+26  &0.03   &APEC+PL  &2.0$\times$10$^{20}$   &2.0 (fix)     &0.67$^{+0.07}_{-0.06}$&(4.5$\pm$0.1)$\times$10$^{-14}$        &$<$1.3$\times$10$^{-14}$              &-&CG$^{c}$     &3\\ 
J115954.66+30  &0.106  &PL         &1.5$\times$10$^{20}$   &2.0 (fix)                  &-           &-         &(4.9$^{+4.9}_{-3.1}$)$\times$10$^{-14}$               &-     &?            &-        \\
J122206.54+13  &0.081  &PL    &3.5$\times$10$^{20}$  &2.0 (fix)          &-                   &-                    &(1.0$^{+1.2}_{-0.7}$)$\times$10$^{-13}$       &-    &Cluster      &5     \\
J125431.43+26  &0.069  &PL    &7.5$\times$10$^{19}$  &1.9$\pm$1.4           &-           &-        &(9.2$^{+1.8}_{-1.6}$)$\times$10$^{-14}$       &4/5   &?            &-        \\
Tol1326-379    &0.028  &PL        &5.5$\times$10$^{20}$  &1.3$\pm$0.4           &-             &-                              &(9.4$^{+1.9}_{-3.1}$)$\times$10$^{-13}$       &- &?            &-  \\
J135908.74+28  &0.073  &APEC+PL &1.3$\times$10$^{20}$   &2.0 (fix)     &0.24 (fix)$^{a}$    &(1.9$^{+0.8}_{-0.7}$)$\times$10$^{-13}$  &(5.0$^{+2.2}_{-2.3}$)$\times$10$^{-14}$  &0.5/2 &Cluster   &1       \\
J153901.66+35  &0.078  &PL      &1.8$\times$10$^{20}$  &2.0 (fix)      &-                   &-            &(1.1$^{+0.7}_{-1.0}$)$\times$10$^{-13}$       &-    &?            &-         \\
J160426.51+17  &0.041	&PL        &3.4$\times$10$^{20}$  &1.1$\pm$0.3&-              &-                   &(1.5$^{+0.5}_{-0.3}$)$\times$10$^{-13}$       &2.5$/$8    & Cluster      &1      \\
J171522.97+57  &0.027  &APEC+PL  &2.2$\times$10$^{20}$  &2.0 (fix)       &1.1$\pm$0.1$^{b}$       &(1.2$^{+0.4}_{-0.5}$)$\times$10$^{-13}$  &$<$1.9$\times$10$^{-14}$     &70/53 & CG$^{c}$   &1     \\
J235744.10-00  &0.076       &  PL            &3.3$\times$10$^{20}$      & 2.0 (fix)         &-                &-           &(9.7$^{+7.9}_{-5.1}$)$\times$10$^{-14}$       &-     &Isolated  &6  \\
\hline\\
\multicolumn{9}{l}{$^{a}$ Due to the poor statistics of the fit this parameter is fixed to the best fit value.}\\    
\multicolumn{9}{l}{$^{b}$ The Abundance parameter is left free and reaches a value of Ab=0.3$^{+0.3}_{-0.1}$.}\\                              
\multicolumn{9}{l}{$^{c}$ CG=compact group of galaxies.}\\
\multicolumn{9}{l}{$^{d}$ (1)-this work, (2)-Koester et al. (2007), (3)-Diaz-Gimenez et al. (2012), (4)-Colbert et al. (2001), (5)-Owen et al. (1995), (6)-Prada et al. (2003).}\\
\end{tabular}
\end{table}
\end{landscape}


\begin{landscape}
\begin{table}
\caption{Radio, optical and X-ray properties of the FR0 sample. (1) Identity number, (2) nuclear radio luminosity at 5~GHz, (3) X-ray nuclear luminosity (2-10~keV) corrected for absorption, (4) [OIII] emission line luminosity. For all luminosities the proper k-correction was considered. (5) Estimated black hole masses for the sources of the sample, (6) Eddington-scaled luminosities ($\dot{L}$=L$_{\rm bol}$/L$_{\rm Edd}$).}
\label{tab4}      
\centering          
\begin{tabular}{l c c c c c }     
\hline\hline       
Source name    &LogL$_{\rm 5~GHz}$ &LogL$_{\rm X,2-10~keV}$    &LogL$_{\rm [OIII]}$$^{a}$   &LogM$_{BH}$ &$\dot{L}$$^{b}$ \\
    &(erg~s$^{-1}$)                &(erg~s$^{-1}$)                       &(erg~s$^{-1}$)        &(M$_{\odot}$)   \\
\hline
J004150.47-09   &40.21				   &41.10		         &39.27        &8.96   	&5.5$\times$10$^{-5}$   \\
J010101.12-00   &39.52		           &$<$41.20	      		&40.39		 &8.43	    &2.4$\times$10$^{-3}$    \\
J011515.78+00   &39.97			       &41.15 			    &39.51        &8.57       &2.3$\times$10$^{-4}$ \\
J015127.10-08   &39.10				   &$<$40.49				&39.29        &7.97   	&5.6$\times$10$^{-4}$    \\
J080624.94+17   &40.69			       &43.20				&39.30    	 &8.39       &2.2$\times$10$^{-4}$   \\
J092405.30+14   &41.38				   &42.16				&40.68	     &9.06   	&1.1$\times$10$^{-3}$  \\
J093346.08+10   &38.46		           &40.17				    &39.14        &8.16    	&2.6$\times$10$^{-4}$ 	\\
J094319.15+36   &40.25				   &40.94			    &39.81        &7.89	    &2.2$\times$10$^{-3}$ 	 \\
J104028.37+09   &39.10                   &39.99				    &39.48        &8.29    	&4.2$\times$10$^{-4}$    \\
J115954.66+30   &39.71			       &$<$40.43 			    &38.48	     &8.97 		&8.7$\times$10$^{-6}$   \\
J114232.84+26   &40.91				   &42.14				&40.24        &8.51    	&1.4$\times$10$^{-3}$ 	\\
J122206.54+13   &40.62		      	   &42.19			    &40.04   	 &8.36	   	&1.3$\times$10$^{-3}$	\\
J125431.43+26   &39.93				   &42.02			    &39.61        &8.59       &2.8$\times$10$^{-4}$  \\
Tol1326-379     &39.79			           &42.29			    &40.60        &8.30 		&5.0$\times$10$^{-3}$   \\
J135908.74+28   &40.51				   &41.69			    &39.48        &8.46   	&2.8$\times$10$^{-4}$ \\
J153901.66+35   &40.82			       &42.20     	        &40.07 	     &8.31   	&1.5$\times$10$^{-3}$ \\
J160426.51+17   &40.37			       &41.64			    &40.02        &8.34    	&1.3$\times$10$^{-3}$	\\
J171522.97+57   &39.46			       &$<$40.16			    &39.46        &8.79   	&1.2$\times$10$^{-4}$  \\
J235744.10-00   &	39.30			       &40.93			    &40.26	     &8.76	&8.5$\times$10$^{-4}$ \\
\hline\\
\multicolumn{6}{l}{$^{a}$ Data are provided by the SDSS Data Release 7 (\ttfamily{http://www.sdss.org/})}\\   
\multicolumn{6}{l}{$^{b}$ $\dot{L}$=L$_{\rm bol}$/L$_{\rm Edd}$. The bolometric luminosity is derived using the relation}     \\                     
\multicolumn{6}{l}{L$_{\rm bol}$=3500~L$_{\rm [OIII]}$ measured by Heckman et al. (2004).} \\
\end{tabular}
\end{table}
\end{landscape}

\section*{Acknowledgements}

The authors thank the anonymous referee for his/her thoughtful comments that helped to improve the paper.
ET acknowledges financial support from ASI-INAF grant 2015-023-R.O. 
This work is based on data from the {\it Chandra}, {\it XMM-Newton} and {\it Swift} Data Archive. Part of this work is based on archival data, software or online services provided by the ASI Science Data Center (ASDC).
We thank the {\it Swift} team for making the ToO observation of Tol1326-379 possible.
Funding for the SDSS and SDSS-II has been provided by the Alfred P. Sloan Foundation, the Participating Institutions, the National Science Foundation, the U.S. Department of Energy, the National Aeronautics and Space Administration, the Japanese Monbukagakusho, the Max Planck Society, and the Higher Education Funding Council for England. The SDSS Web Site is http://www.sdss.org/.







\appendix

\section{Notes and analysis details on individual sources}

{\bf J004150.47-09}\\ 
At the centre of the cluster Abell~85. This cluster presents X-ray cavities created by the central AGN, as it is evident from the {\it Chandra} 3-7~keV image (Hlavacek-Larrondo et al. 2013). The presence of such cavities suggests that this AGN produced in the past an extended radio source able to excavate the external medium; this is an indication of recurrency. \\

\noindent
{\bf J010101.12-00}\\
This source is part of the B15 sample (their ID590) observed with the JVLA. 
For this object we could only obtain an upper limit on the 2-10~keV flux estimated from the \emph{Chandra} count rate using WebPIMMS and assuming a spectral slope $\Gamma$=2.    \\

\noindent
{\bf J011515.78+00}\\ 
This source, GIN061, is part of the sample of FR0s presented in B15 (their ID605). It lies at the outskirt of the Abell cluster A168 (Dressler 1980), as shown by the {\it Chandra} (Yang et al. 2004) and {\it XMM-Newton} (Figure~3) X-ray images. The good quality of the XMM-Newton spectrum allowed us to constrain the parameters of the power-law ($\Gamma$=2.0$\pm$0.2) and the thermal component (kT=0.8$\pm$0.1). \\

\noindent
{\bf J015127.10-08}\\ 
The X-ray spectrum of this source (a.k.a. NGC~0707) can be reproduced by a power-law plus a thermal model with kT$\sim$0.2~keV. This latter component is dominant, and for the power-law we could estimate only an upper limit on the 2-10~keV flux.\\

\noindent
{\bf J080624.94+17}\\ 
From the work of Koester et al. (2007) the source seems to reside in a galaxy cluster. However, the statistics of the X-ray data is too low to constrain the parameters of a thermal component. Therefore the spectrum is fitted with an absorbed power--law fixing $\Gamma$=2. \\

\noindent
{\bf J092405.30+14}\\ 
The source resides close to the centre of the cluster Abell~795.  The X-ray spectrum can be well reproduced by a power-law with $\Gamma$=1.8 plus an {\small APEC} component  with kT=1.3~keV related to the thermal gas of cluster.    \\

\noindent
{\bf J093346.08+10}\\ 
In NGC~2911 an extended jet radio emission on subarcsecond scales has been observed (Mezcua \& Prieto 2014), meaning that this source is powerful enough to launch radio jets, at least on pc scales.\\

\noindent
{\bf J094319.15+36}\\ 
NGC~2965. The \emph{Swift} spectrum of this source can be modeled with a steep power-law ($\Gamma$=2.3) only absorbed by Galactic column density. However, the low statistics prevented us from establishing the presence of a possible thermal component.\\

\noindent
{\bf J104028.37+09}\\ 
NGC~3332 resides in an isolated elliptical galaxy (Colbert et al. 2001). Its X-ray spectrum can be reproduced by a steep ($\Gamma$=2.2$\pm$0.4) power-law absorbed only by the Galactic column density.\\

\noindent
{\bf J114232.84+26}\\ 
NGC~3826 is a member of a compact group of galaxies included in the photometric catalog of Diaz-Gimenez et al. (2012).
The source lies at the edge of the XMM-Newton/PN field of view (offset=13.7 arcmin) since the main target was the star GJ436. Inspite of the poor statistics, the spectrum is better fitted with two components: a thermal model, probably related to the ICM and a power--law emerging above 2~keV. When the spectral slope, initially fixed to $\Gamma$=2, was let free to vary it assumed a smaller value suggesting the presence of intrinsic absorption.  We could not constrain the value of the intrinsic N$_{\rm H}$ but providing an upper limit on this parameter N$_{\rm H}<$3.4$\times$10$^{23}$~cm$^{-2}$.\\

\noindent
{\bf J115954.66+30}\\ 
This source at z=0.106 is among the three farthest objects of the sample. We do not have any information about the environment 
either from the {\it Swift} X-ray image or from the literature. We reproduced the XRT spectrum with a power-law fixing the photon index to $\Gamma$=2.\\

\noindent
{\bf J122206.54+13}\\ 
The source is also known with the name VPC0184 and is associated to the cluster Abell~1526, that however is not visible in our \emph{Swift}/XRT image. 
The X-ray spectrum can be reproduced by a power-law with photon index fixed to a value of 2. Indeed, the statistics  is not good enough to constrain this parameter nor to establish the possible presence of thermal gas related to the cluster. For this reason we are aware that the X-ray flux that we report in Table~3 could be overestimated. \\

\noindent
{\bf J125431.43+26}\\ 
This source is already present in the Chandra Source Catalog to Sloan Digital Sky Survey sample (CSC/SDSS) by Trichas et al. (2013).
The X-ray spectral parameters presented in this paper are in agreement with those found by Trichas et al. We have no information on the environment either from the X-ray image nor from the literature. \\

\noindent
{\bf Tol1326-379}\\
Tol1326-379 is a FR0 source at z=0.0284 hosted in an early type galaxy. It is the first FR0 with an associated $\gamma$-ray counterpart (Grandi et al. 2106). Indeed, it is listed in the 3LAC catalog as 3FGLJ1330.0-3818 (Ackermann et al. 2015). As it is evident from Table~3 the photon index of this gamma-ray FR0 is quite flat $\Gamma$=1.3$\pm$0.4. However, the inclusion of an intrinsic absorber did not significantly improve the fit. We could only estimate an upper limit to the intrinsic column density N$_{\rm H}$<3.4$\times$10$^{22}$~cm$^{-2}$.\\

\noindent
{\bf J135908.74+28}\\ 
This source, Zw162-39, resides in the Abell cluster A1831 (Feretti \& Giovannini 1994), as it shown in Figure~3. Indeed, the residuals in the X-ray spectrum suggest the presence of a thermal component in addition to the power-law. The estimated best-fit value of the gas temperature is kT=0.24~keV. \\

\noindent
{\bf J153901.66+35}\\  
The very low statistics of the X-ray spectrum prevents us from establishing the presence of diffuse emission around the source.
Therefore the spectrum is fitted with an absorbed power-law fixing $\Gamma$=2. \\

\noindent
{\bf J160426.51+17}\\ 
This source, NGC~6040B, lies at the outskirt of the cluster Abell~2151 (a.k.a. Hercules cluster). Together with NGC~6040A, located at 0.5~arcmin, it forms an interacting pair (de Vaucoulers et al. 1976). The X-ray spectrum is well reproduced by an absorbed power-law. The flatness of the photon index ($\Gamma$=1.1, see Table~3) suggests the possible presence of extra absorption. However, the low statistics prevent us from constraining the value of this additional component, for which we could only estimate an upper limit N$_{\rm H}$<4.9$\times$10$^{22}$~cm$^{-2}$.\\

\noindent
{\bf J171522.97+57}\\
NGC~6338 is part of a cool-core cluster (Bharadwaj et al. 2015). The X-ray spectrum is dominated by the thermal component of the cluster, for which we could constrain also the abundances to a value of 0.3. The power--law is marginally significant in fact we could only determine an upper limit to the flux of this component.\\

\noindent
{\bf J235744.10-00}\\ 
The X-ray flux of this source, that is part of the B15 sample (their ID535), is provided by the 3XMM-DR6 catalog (Rosen et al. 2016). 
Prada et al. (2003) lists this source as an isolated galaxy.\\

\begin{table}
\caption{Details of the data reduction for the sample of FR0s.}
\centering          
\begin{tabular}{l l l l l l l}     
\hline\hline       
{\bf Source}   &{\bf Reg.}       &{\bf Shape$^{a}$}   	    &{\bf R$^{b}$}   &{\bf R$_{in}$$^{c}$}        &{\bf R$_{out}$$^{d}$}    &{\bf X-ray}   \\
     {\bf (Instr.)}            &                    &                 &$['']$       &$['']$     &$['']$      & {\bf cluster$^{e}$} \\
\hline
J004150.47-09       &s         &c           &2.8               &-              &-                &Yes\\
{\tiny (Chandra)}   &b         &a            &-               &5.5                            &7.8              &\\
\hline
J010101.12-00       &s         &c					&5					&-				&-				&?\\
{\tiny (Chandra)}	         &b           &a              &-                       &7.1                          &9.6              &\\
\hline
J011515.78+00       &s           &c                     & 16                 &-                           &-                   &Yes\\
{\tiny (XMM)}               &b           &c                     & 16                 &-&-&\\
\hline
J015127.10-08       &s             &c                      &20                  &-                          &-                  &?\\
{\tiny (Swift)}              &b              &a                   &-                    &40                       &80                   &\\     
\hline
J080624.94+17        &s            &c                      &20                     &-                         &-                  &?\\
{\tiny (Swift)}              &b           &a                   &-                       &40                     &80                      &\\
\hline
J092405.30+14         &s         &c                        &2.5                  &-                         &-                 &Yes\\
{\tiny (Chandra)}             &b             &a                   &-                     &3.8                     &5.4                 &\\
\hline
J093346.08+10        &s		   &c                       &20   			&-				    &-                  &?\\
{\tiny (Swift)}             &b             &a                   &-                      &40                      &80                   &\\
\hline
J094319.15+36        &s 	            &c                        &20                &-                          &-         	&?\\
{\tiny (Swift)}              &b            &a                    &-                    &40                      &80                 &\\
\hline 
J104028.37+09        &s                   &c                   &22                 &-                         &-             &No\\               
{\tiny (XMM)}                &b          &c                        &22                 &-                         &-                 & \\
\hline
J114232.84+26        &s             &c                 &26.5                     &-                         &-               &?\\ 
{\tiny (XMM)}              &b           &a                     &-                         &57.4                     &87.7           &\\
\hline
J115954.66+30       &s		&c				&20					&-		&-				&?\\
{\tiny (Swift)} 		   &b			&a			&-			&40					&80			&\\
\hline
J122206.54+13      &s           &c                      &20                     &-                         &-         &?\\
{\tiny (Swift)}            &b              &a                 &-                         &40                     &80            &\\
\hline
J125431.43+26       &s     &c                       &2.7                       &-                          &-            &No\\
{\tiny (Chandra)}           &b            &a                  &-                   &4.5                       & 8.8                   &\\
\hline
Tol1326-379         &s          &c                    &20            &-          &-                    &No\\
{\tiny (Swift)}		    &b      &a                &-               &40      &80                  &\\
\hline
J135908.74+28       &s     &c                      &4.5                    &-                         &-                   &Yes\\
{\tiny (Chandra)}           &b             &a                  &-                        &8.1                     &16.7             &\\
\hline
J153901.66+35       &s                 &c                        &20                 &-                         &-         &?   \\
{\tiny (Swift)}             &b               &a                    &-                    &40                      &80              &\\
\hline
J160426.51+17	     &s     &c                        &6.6                &-                         &-                   &Yes\\
{\tiny (Chandra)}	          &b          &a                    &-                     &8.7                    &14.5                  &\\
\hline
J171522.97+57         &s       &c                       &3.3                 &-                         &-                &Yes\\
{\tiny (Chandra)}                      &b   &a                   &-                     &4.7                     &9.5                        &\\
\hline
J235744.10-00$^{f}$    &-			&-				&-					&-								&-					&?\\   
\hline
\hline\\

\multicolumn{7}{l}{$^{a}$ Shape of the extraction region: c=circle, a=annulus.}\\
\multicolumn{7}{l}{$^{b}$ Radius of the circular region.}\\
\multicolumn{7}{l}{$^{c}$ Internal radius of the annular region.}\\
\multicolumn{7}{l}{$^{d}$ External radius of the annular region.}\\
\multicolumn{7}{l}{$^{e}$ The source is within (or at the edge of) a cluster or a group of}\\ 
\multicolumn{7}{l}{galaxies in the X-ray image.}\\                              
\multicolumn{7}{l}{$^{f}$ The source is part of the 3XMM-DR6 catalog of serendipitous }\\
\multicolumn{7}{l}{sources (Rosen et al. 2016).}\\
\end{tabular}
\end{table}

\section{FRI}
Radio, optical and X-ray data of the 35 3CR-3CRR/FRIs in Table~B1 are from: Buttiglione et al. (2009, 2010, 2011), Leipski et al. (2009), Lanz et al. (2015), Balmaverde et al. (2006), Hardcastle et al. (2009), Ogle et al. (2010), Migliori et al. (2011), Mingo et al. (2014), Dasadia et al. (2016). 

For eleven sources the X-ray luminosities were obtained directly analyzing {\it Chandra} data using the software CIAO (v.4.7) and calibration database v.4.6.9. We followed standard procedures to extract source and background spectra. For the source we chose a circular region varying between 2.5$"$ and 7$"$. Only in two cases, e.g. 3C~371 and 3C~465, data were strongly piled up and therefore the spectra were extracted from annular regions (r$_{in}$=0.5$"$, r$_{out}$=2.5$"$).
All background spectra were taken from annular regions with r$_{in}$ and r$_{out}$ ranging between 3$"$-16$"$ and 4$"$-30$"$, respectively. Data were then grouped to a minimum of 15 counts per bin over the energy range 0.5-7~keV. This allows us to apply the $\chi^2$ statistics. For 3C~29, 3C~76.1 and 3C~129 data were not grouped and the C-statistics was applied.
As a baseline model we adopted a power-law absorbed by Galactic column density. The photon index was let free to vary except for 3C~310 and 3C~424: in these cases $\Gamma$ was fixed to 2. When the residuals were not satisfactory a thermal component was added to the data. Finally, in case the value of the photon index was very flat, we included an intrinsic N$_{H}$ to the model. The results of the spectral analysis are summarized in Table~B2.


\begin{landscape}
\begin{table}
\caption{Multi-wavelength properties of the comparison sample of FRIs: (1) name, (2) redshift, (3) radio core luminosity at 5~GHz, (4) X-ray (2-10~keV) luminosity, (5) [OIII] emission line luminosity, (6) NVSS 1.4~GHz radio luminosity, (7) black hole masses from Marchesini et al. (2004), Woo \& Urry (2002), Graham et al. (2001), Cao \& Rawlings (2004), (8) Eddington-scaled luminosities ($\dot{L}$), (9) references for the X-ray luminosities. For all  luminosities the proper k-correction was considered.}
\centering
\begin{tabular}{|l|r|r|r|r|r|r|r|r|r|r|r|r|r|r|r|r|r|r|r|r|r|r|r}
\hline
  \multicolumn{1}{l|}{Name} &
  \multicolumn{1}{c|}{z} &
  \multicolumn{1}{c|}{Log L$_{\rm (5~GHz)}$} &
  \multicolumn{1}{c|}{Log L$_{\rm (2-10~keV)}$} &
  \multicolumn{1}{c|}{Log L$_{\rm [OIII]}$} &
  \multicolumn{1}{l|}{Log L$_{\rm (1.4~GHz)}$} &
  \multicolumn{1}{l}{M$_{\rm BH}$} &
  \multicolumn{1}{c}{$\dot{L}$$^{a}$} &
  \multicolumn{1}{r|}{Ref.}\\
\hline
3C~29   & 0.0448 & 40.33 & 41.38  & 40.09   &41.32 	    &8.81   &5.1$\times$10$^{-4}$	 &2\\
3C~31 	& 0.0169 & 39.4 & 40.73   & 39.46 	& 39.90 	&7.89   &1.0$\times$10$^{-3}$	 &1\\
3C~66B 	& 0.0215 & 39.9 & 41.17   & 40.05  	& 40.32 	&8.84   &4.4$\times$10$^{-4}$	 &1\\
3C~75 	& 0.0232 & 39.3 & <41.11  & <39.92  & 37.86 	&9.0    &<2.2$\times$10$^{-4}$    &1\\
3C~76.1 & 0.0324 & 39.07 & 40.92  & 39.85  	& 40.74 	&8.13   &1.4$\times$10$^{-3}$    &2\\
3C~78 	& 0.0288 & 40.88 & 42.37  & 39.41 	& 41.17 	&8.98   &7.2$\times$10$^{-5}$    &1\\
3C~83.1 & 0.0251 & 39.11 & 41.2   & <39.5   & 40.40	    &9.01   &<8.3$\times$10$^{-5}$   &1\\
3C~84  	& 0.0176 & 42.08 & 42.7   & 41.6  	& 41.33 	&9.28   &5.6$\times$10$^{-3}$    &1\\
3C~89  	& 0.138  &41.08 &41.77    & 40.51   & 42.09     &8.83	&1.3$\times$10$^{-3}$	 &5\\
3C~129 	& 0.0208 & 39.21 & 40.11  & <39.85 	& 40.74 	& -     & -                      &2\\
3C~130 	& 0.109 & 40.64 & 41.57   & - 		& 41.62 	& -     & -                 	 &2\\
3C~189 	& 0.043 & 40.53 & 41.94   & 39.94 	& 41.20 	&8.93   &2.7$\times$10$^{-4}$    &1\\
3C~264 	& 0.0217 & 39.98 & 41.87  & 39.2  	& 40.71 	&8.85   &6.0$\times$10$^{-5}$    &3\\
3C~270 	& 0.0074 & 39.2 & 40.73   & 38.96 	& 39.83 	&8.57   &6.6$\times$10$^{-5}$    &1\\
3C~272.1 & 0.0037 & 38.36 & 39.51 & 38.2    & 39.12 	&8.35   &1.9$\times$10$^{-5}$    &1\\
3C~274 	& 0.0037 & 39.72 & 40.35  & 38.99 	& 40.76 	&8.26   &1.4$\times$10$^{-4}$    &1\\
3C~293 	& 0.0452 & 40.36 & 42.78  & 39.8  	& 41.42 	&7.99   &1.7$\times$10$^{-3}$    &3\\
3C~296 	& 0.0237 & 39.61 & 41.38  & 39.78  	& 40.39 	&9.13   &1.2$\times$10$^{-4}$    &1\\
3C~310 	& 0.053  & 40.41 & <40.41 & 40.05 	& 41.86 	&8.29   &1.5$\times$10$^{-3}$	 &2\\
3C~315 	& 0.1083 & 41.34 & 41.68  & 40.86 	& 42.17 	&8.7    &3.9$\times$10$^{-3}$    &4\\
3C~317 	& 0.0342 & 40.64 & 41.44  & 40.35 	& 41.30 	&8.80   &9.5$\times$10$^{-4}$	 &1\\
3C~338 	& 0.0303 & 39.96 & <40.76 & 39.57	& 41.02 	&9.23   &5.9$\times$10$^{-5}$    &1\\
3C~346 	& 0.162  & 41.74 & 43.45  & 41.32 	& 42.53 	&8.89   &7.2$\times$10$^{-3}$    &1\\
3C~348 	& 0.154  & 40.36 & <42.69 & 40.4  	& 41.85 	&8.84   &9.8$\times$10$^{-4}$    &1\\
3C~371 	& 0.05   & 41.55 & 42.89  & 40.94 	& 41.17 	& -     & -                      &2\\
3C~386 	& 0.017  & 38.65 & 39.75  & 40.2 	& 40.62 	&8.5    &1.3$\times$10$^{-3}$    &4\\
3C~402 	& 0.0259 & 39.48 & 41.20  &<39.42 	& 40.60 	&8.18   &<4.7$\times$10$^{-4}$   &2\\
3C~424 	& 0.126	 & 40.56 & 42.57  & 40.80 	& 42.15	    &8.17   &1.1$\times$10$^{-2}$    &2\\
3C~438 	& 0.29   & 41.15 & <43.14 & 41.46 	& 43.01 	&8.80   &1.2$\times$10$^{-2}$    &3\\
3C~442  & 0.0263 & 38.19 & 41.17  & 39.21   & 40.22     &8.28   &2.3$\times$10$^{-4}$    &2\\
3C~449 	& 0.0181 & 39.04 & <40.46 & 39.19   & 39.32     &7.71   &8.1$\times$10$^{-4}$    &1\\
3C~465 	& 0.0303 & 40.44 & 40.60  & 39.81   & 40.43     &9.32   &8.3$\times$10$^{-5}$    &2\\
NGC~6109 & 0.0296 & 39.44 & 40.06 & -       & 40.01	    & -     & -  		             &3\\
NGC~6251 & 0.024 & 40.35 & 42.76  & 39.86   & 40.15     &8.77   &3.3$\times$10$^{-4}$    &6\\
NGC~7385 & 0.026 & 39.90 & 41.07  & -       & 40.41     & -     &  -                     &2\\
\hline
\multicolumn{7}{l}{$^{a}$ $\dot{L}$=L$_{\rm bol}$/L$_{\rm Edd}$. The bolometric luminosity is derived from the relation L$_{\rm bol}$=3500~L$_{\rm [OIII]}$}\\
\multicolumn{7}{l}{(Heckman et al. 2004) as for FR0s (see Table~4).}\\
\multicolumn{7}{l}{(1)-Balmaverde et al. (2006), (2)-this work, (3)-Hardcastle et al. (2009), (4)-Ogle et al. (2010),}\\
\multicolumn{7}{l}{(5)-Dasadia et al. (2016), (6)-Migliori et al. (2011).}\\
\end{tabular}
\end{table}
\end{landscape}



\begin{landscape}
\begin{table}
\caption{Spectral results for the eleven FRIs analyzed with {\it Chandra}.}
\centering
\begin{tabular}{|l|c|c|c|c|c|c|c|r|r|r|r|r|r|r|r|r|r|r|r|r|r|r|r}
\hline
  \multicolumn{1}{l|}{Name} &
  \multicolumn{1}{c|}{N$_{\rm H, Gal}$} &
  \multicolumn{1}{c|}{N$_{\rm H, int}$} &
  \multicolumn{1}{c|}{$\Gamma$} &
  \multicolumn{1}{c|}{kT} &
  \multicolumn{1}{c|}{F$_{\rm th, 0.5-5~keV}$} &
   \multicolumn{1}{c|}{F$_{\rm nucl, 2-10~keV}$}\\
        &{\tiny (cm$^{-2}$)}   &{\tiny (cm$^{-2}$)}   &   & {\tiny (keV)}     &{\tiny (erg~cm$^{-2}$~s$^{-1}$)}   &{\tiny (erg~cm$^{-2}$~s$^{-1}$)}\\
\hline
3C~29      & 3.7$\times$10$^{20}$ & <0.2$\times$10$^{23}$   & 2.4$\pm$0.5 	       & -    & -     &(5.0$^{+2.5}_{-1.7}$)$\times$10$^{-14}$\\
3C~76.1    & 9.5$\times$10$^{20}$ & -                       &1.9$^{+1.2}_{-0.9}$  & -  	& - 	&(1.5$^{+0.9}_{-0.6}$)$\times$10$^{-14}$\\
3C~129 	& 5.9$\times$10$^{21}$ & -                       &2.3$^{+0.8}_{-0.7}$   & - 	& -	&(1.3$^{+0.9}_{-0.6}$)$\times$10$^{-14}$\\
3C~130 	& 3.7$\times$10$^{21}$ & -                       &1.9$\pm$0.4 		 &- 		& -  &(2.4$\pm$0.7)$\times$10$^{-14}$\\
3C~310     & 3.7$\times$10$^{20}$ & -         &2.0 (f)  &1.2$^{+0.2}_{-0.5}$ &(6.4$^{+4.1}_{-4.3}$)$\times$10$^{-15}$ &<2.4$\times$10$^{-15}$\\
3C~371 	& 4.2$\times$10$^{20}$ & -         &1.3$\pm$0.06 	& 0.3$^{+0.2}_{-0.1}$ &(2.7$^{+1.8}_{-1.7}$)$\times$10$^{-14}$ 	&(1.4$\pm$0.1)$\times$10$^{-12}$\\
3C~402 	& 1.1$\times$10$^{21}$ & -        & 2.1$\pm$0.4  &-   	& - 	&(1.0$\pm$0.2)$\times$10$^{-13}$\\
3C~424 	& 7.0$\times$10$^{20}$  & -        & 2.0 (f) 	& -  & -   	    &(8.6$^{+2.4}_{-2.0}$)$\times$10$^{-14}$\\
3C~442     & 4.9$\times$10$^{20}$  & (0.9$\pm$0.3)$\times$10$^{22}$ & 1.9$\pm$0.3  & -   & -     &(9.9$^{+20.1}_{-6.2}$)$\times$10$^{-14}$\\
3C~465 	& 1.9$\times$10$^{21}$  & -         & 1.7$^{+0.8}_{-1.1}$  & 0.9$\pm$0.05   &(8.6$^{+1.3}_{-1.4}$)$\times$10$^{-14}$  	&(2.2$^{+2.5}_{-1.6}$)$\times$10$^{-14}$\\
NGC~7385   & 1.6$\times$10$^{21}$  & -         & 2.6$\pm$0.3  & -      & -     &(7.6$^{+2.1}_{-1.6}$)$\times$10$^{-14}$\\
\hline
\end{tabular}
\end{table}
\end{landscape}


\bsp	
\label{lastpage}
\end{document}